\documentclass[aps,prl,twocolumn, superscriptaddress]{revtex4-1}
\usepackage{graphicx}
\usepackage{dcolumn}
\usepackage{bm}
\usepackage{graphicx}
\usepackage{dcolumn}
\usepackage{bm}
\usepackage{subfigure}
\usepackage{color}
\usepackage{amsmath}

\begin{document}

\newcommand{\change}[1]{\textcolor{red}{#1}} 
\newcommand{\BU}{Department of Mechanical Engineering, Boston University, Boston, Massachusetts 02215, USA}
\newcommand{\Bilkent}{Department of Mechanical Engineering, Bilkent University, Ankara 06800, Turkey}


\title{Generalized Knudsen Number for Unsteady Fluid Flow}

\author{V. Kara}
\affiliation{\BU}

\author{V. Yakhot}
\affiliation{\BU}

\author{K. L. Ekinci}
\email[Electronic mail: ]{ekinci@bu.edu}
\affiliation{\BU}

\date{\today}

\begin{abstract}

We explore the scaling behavior of an unsteady flow that is generated by an oscillating  body of finite size in a gas. If the gas is gradually rarefied, the Navier-Stokes equations begin to fail and a kinetic description of the flow becomes more appropriate. The failure of the Navier-Stokes equations can be thought to take place via two different physical mechanisms: either the continuum hypothesis breaks down as a result of a finite size effect; or local equilibrium is violated due to the high rate of strain. By independently tuning the relevant linear dimension and the frequency of the  oscillating body, we can experimentally observe these two different physical mechanisms. All the experimental data, however, can be collapsed using a single dimensionless scaling parameter that combines the relevant linear dimension and the frequency of the body. This proposed Knudsen number for an unsteady flow  is rooted in a fundamental symmetry principle, namely Galilean invariance.

\end{abstract}

\maketitle

The Navier-Stokes (NS) equations of hydrodynamics can be  obtained perturbatively from the kinetic theory of gases  in the limit of small Knudsen number,  ${\rm Kn} = \frac{\lambda}{\cal L} \to 0$ \cite{Landau_Physical_Kinetics}. Here, $\lambda$ is the mean free path in the gas, and $\cal L$ represents a characteristic length scale of the flow. As ${\rm Kn} \to 0$,  it follows from statistical mechanics  that  density fluctuations in the gas vanish \cite{Huang}, leading to the notion of a ``fluid particle." This continuum hypothesis becomes less accurate as $\rm Kn$ grows, eventually leading to the failure of the NS equations for $\rm Kn \gtrsim 0.1$. Likewise, the NS equations  break down  if the local value of the strain rate, $S_{ij}=\frac{1}{2} \left( \frac{\partial u_{i}}{\partial x_{j}}+ \frac{\partial u_{j}}{\partial x_{i}} \right)$, becomes so large  that the condition $\tau S_{ij} \ll 1$  no longer holds. Here, $u_i$ represents the velocity vector, and $\tau$ is the relaxation time that characterizes the rate of decay of a  perturbation  to thermodynamic equilibrium. As $\tau S_{ij}$ grows, the fluid particle becomes deformed on  shorter and shorter time scales, eventually violating the local equilibrium assumption. For a broad class of flows,  breakdown of the continuum hypothesis and  violation of local equilibrium can be thought to be equivalent, because  $\tau S_{ij}\sim \tau \frac {U}{\cal L}\sim \frac{\lambda}{c}\frac{U}{\cal L}\sim \rm Ma \times Kn$. Here, the Mach number ${\rm Ma}=\frac {U}{c}$ compares  the speed of sound $c$  to the characteristic flow velocity $U$, and it is assumed to remain small and slowly varying. Thus, either $\rm Kn$ or $\tau S_{ij}$ emerges as the relevant scaling parameter for determining the crossover from hydrodynamics to kinetic theory.

To demonstrate the limitations of the above-described widely-accepted reasoning, we consider the canonical problem of an \emph{infinite} plate oscillating at a ${\it prescribed}$  angular frequency $\omega_0$ in a gas (Stokes Second Problem) \cite{Landau_Fluids}. We assume the oscillation amplitude to be small and the geometry  to be such that  the velocity field is  $u_x(x,y,0)=U_0 {\cos \omega_0 t}$, $u_{y}=0$, and $u_{z}=0$. Since  the plate is infinite ($l \to \infty$), the ``standard" size-based Knudsen number ${\rm Kn}_l = \frac{\lambda}{l}$  remains zero at all limits and cannot be relevant. The  scaling parameter here is the Weissenberg number, $\rm Wi=\omega_0 \tau$ \cite{Ekinci, Rarefied}, and one can recover the correct Knudsen number, $\rm {Kn}_\delta=\frac{\lambda}{\delta}$, using the boundary layer thickness, $\delta=\sqrt{\frac{2\nu_g}{\omega_0}}$. (Indeed, $\rm {Kn}_\delta \sim \sqrt {\rm Wi}$, given the kinematic viscosity is  $\nu_g \sim \frac{\lambda^2}{\tau}$.) Regardless, $\tau S_{ij} \approx \tau \frac {U_0}{\delta} \sim {\rm Ma \times Kn}_\delta$. Thus, as above, the validity of the NS equations (and the scaling properties of the flow) is determined either by the flow length scale  (${\rm Kn}_\delta$) or by the flow time scale ($\tau S_{ij}$ or $\rm Wi$), and  both  parameters lead to the same conclusion. While this analysis for an infinite plate is reasonable, it does not work for a finite plate (or a finite-sized body). For a finite-sized body,  ${\rm Kn}_l$ may be non-zero at some limit and appear in the problem alongside $\rm Wi$. This is because the oscillation frequency $\omega_0$ is in general independent of the linear dimensions of the body and  an externally-prescribed parameter. Recent literature on  scaling of such flows reflects this complexity: some reports suggest    ${\rm Kn}_l$ scaling \cite{Bullard, Martin, Bhiladvala} and others  $\rm Wi$ scaling \cite{Karabacak, Ekinci,Svitelsky}. The purpose of the present  work is to study this non-trivial limit and to recover, both experimentally and theoretically, the  universal scaling hidden in the apparent contradictions.

\begin{figure}
\includegraphics[width=3.375 in]{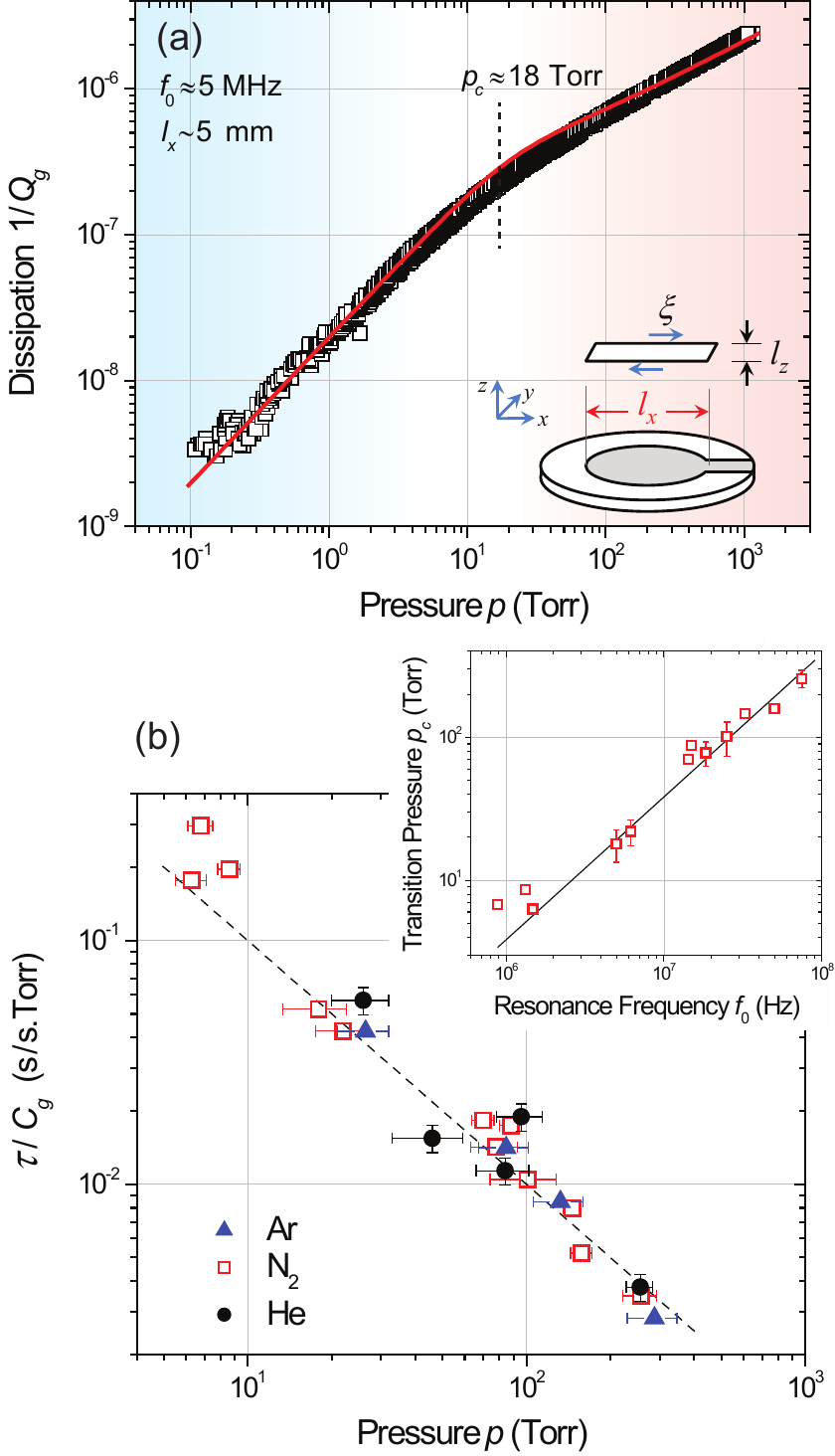}
\caption{(a) Dissipation in N$_2$ as a function of pressure for a quartz crystal (inset) oscillating in shear mode at $f_0 \approx 5$ MHz. Solid line is a fit to Eq.~(\ref{eq:plate_soln}). Transition from   the kinetic to viscous regime  occurs at  $p_c \approx 18$ Torr.  (b)  The inset shows  $p_c$ \emph{vs.}   $f_0$ for different quartz crystals in N$_2$  (${\rm Kn}_l \approx 0$). The linear fit gives the empirical $\tau$ as a function of $p$. The main figure shows $\tau / {\cal C}_g$ for He, N$_2$, and Ar as a function of $p$. Normalization by ${\cal C}_g$  accounts for the  differences between gases \cite{Supplemental}. Dashed line is $1/p$. Error bars are not shown when smaller than symbols.}
\label{fig:tauP2}
\end{figure}

Our experimental measurements are based on  quartz crystals, and micro- and nano-mechanical resonators. When driven to oscillations  in a gas, these structures generate oscillatory flows and dissipate energy. The gases  used are  high-purity He, N$_{2}$, and Ar. The approximate equation of motion of a mechanical resonator (in any resonant mode)  is that of a damped harmonic oscillator: $\ddot \xi + \frac {\omega_0}{Q_t} \dot \xi + {\omega_0}^2 \xi = {\cal F}(t)/m_r$, where $\xi(t)$ is the amplitude, $m_r$ is the mass, $1 \over Q_t$ is the total (dimensionless)  dissipation, and $\omega_0= 2 \pi f_0$ is the angular frequency of the mode driven by the sinusoidal force ${\cal F}(t)$.  In a typical experiment, the pressure $p$ of the gas is changed, and $1 \over Q_t$ and $\omega_0$ are measured. For all practical purposes, $\omega_0$ stays constant through  $p$ sweeps. To obtain the (dimensionless) gas dissipation $\frac {1} {Q_g}$, we calculate $\frac {1} {Q_g} = \frac {1} {Q_t} - \frac {1} {Q_0}$, where $\frac {1} {Q_0}$ is the intrinsic dissipation  (obtained at the lowest $p$). Relevant parameters of  our resonators and other details can be found in the Supplemental Material \cite{Supplemental}.

All our ${1\over Q_g}$ \emph{vs}. $p$ data possess similar features (Figs. 1a, 2a, 3a, 3b S2-S10).  At  low $p$, ${1\over Q_g} \propto p$. This is the kinetic limit \cite{Christian, Gombosi}, where the  mean free path $\lambda$ and the relaxation time $\tau$ of the gas are both large. At high $p$, the NS equations are to be used \cite{Landau_Fluids}. The crossover between these two asymptotes (transitional flow regime) manifests itself as a slope change in the data. The pressure $p_c$, around which this transition occurs, is therefore a fundamentally important parameter and  provides insight into how this flow scales. ($p_c, \tau_c$ and $\lambda_c$ henceforth indicate transition values.)

\begin{figure*}
\centering
\includegraphics[width=6.75in]{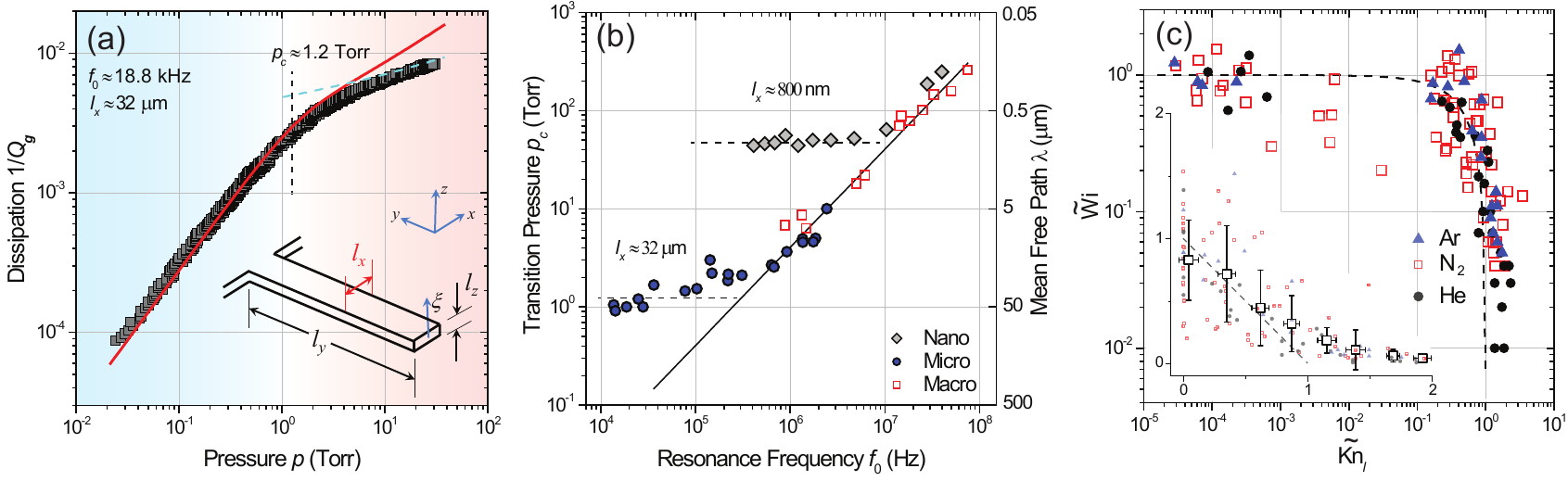}
\caption{(a) Dissipation \emph{vs.} pressure for a microcantilever (inset) with $l_x \times l_y \times l_z \approx 32 \times 350 \times 1~\mu \rm m^3$ and $f_0 \approx 18.8$ kHz in N$_2$. Solid line is a fit to Eq.~(\ref{eq:soln_finite_plate});  dotted (blue) line is a fit to the cylinder solution; $p_c\approx 1.2$ Torr. (b)  $p_c$ \emph{vs.}   $f_0$ in N$_2$  for three sets of devices with different characteristic dimensions. Diamonds are nanocantilevers  from ref. \cite{Kara};  circles are  microcantilevers; squares are macroscopic resonators from Fig. 1b. (c) $\tilde {\rm Wi}$ and  ${\tilde {\rm Kn}}_l$ in He, N$_2$, and Ar for all  devices. Dashed line is  $ \tilde {\rm Wi} + {\tilde {\rm Kn}}_l = 1$. The inset shows the same data using linear axes; the large data points correspond to binned average values.}
\label{fig:Qgas}
\end{figure*}

We first analyze the dissipation  of a macroscopic quartz crystal resonator in shear-mode oscillations in N$_2$ (Fig. 1a). The resonance frequency is $f_0 = \frac {\omega_0} {2 \pi} \approx 5$ MHz, and the relevant linear dimension is \emph{roughly} the diameter of the metal electrode on the quartz, $l_x \sim 5$ mm (Fig. 1a inset). For the shown  pressures, ${\rm Kn}_l = \frac {\lambda}  {l_x}$ is  in the range  $10^{-5} \lesssim {\rm Kn}_l  \lesssim 10^{-1}$,  found using $\lambda \approx \frac{{{k_B}T}}{{\sqrt 2 \pi {{d_g}^2}p}}$, where $k_BT$ is the thermal energy and $d_g$ is the diameter of a N$_2$ molecule. Because  ${\rm Kn}_l$ remains small,  we treat the quartz as an infinite plate  and ${\rm Wi}= \omega_0 \tau$ is left as the only relevant scaling parameter. The transition from molecular flow ($\omega_0 \tau \gg 1$) to viscous flow ($\omega_0 \tau \ll 1$) must take place  at ${\tilde {\rm Wi}}=\omega_0 \tau_c \approx 1$. Hence, we call this the ``high-frequency limit." Next, we perform the same $\frac {1} {Q_g}$ \emph{vs.} $p$ measurement  on similarly large quartz resonators but with different $f_0$. We  determine $p_c$ consistently for all  by finding the pressure at which  $\frac {1} {Q_g}$ deviates  from the low-$p$ asymptote by  $25\%$. The inset of Fig. 1b shows the measured $p_c$ values in N$_2$ as a function of  $f_0$. The data scale as ${{p_c}} = {\rm {constant}} \times f_0$. This is consistent with the flow being scaled by ${\rm Wi}= \omega_0 \tau$ and $\omega_0 \tau_c \approx 1$ determining the transition: $\tau = \frac{ {\cal C}_{\rm N_2}}{ p}$ for a near-ideal gas with  ${\cal C}_{\rm N_2}$ being a constant;  $\omega_0 \tau_c \approx \omega_0 \frac {{\cal C}_{\rm N_2}} {p_c} \approx 1$, and ${p_c} \approx 2\pi {{\cal C}_{\rm N_2}} \times f_0$. The experiment  provides the \emph{empirical} value ${\cal C}_{\rm N_2}= 610 \pm 30 \times 10^{-9}$  in units of s$\cdot$Torr. Repeating the same experiment for He and Ar, we find ${\cal C}_{\rm He}= 560 \pm 70 \times 10^{-9}$  and ${\cal C}_{\rm Ar}= 750 \pm 80 \times 10^{-9}$, both in units of s$\cdot$Torr.  Figure 1b (main) is a collapse plot of $\tau/{\cal C}_g$ for all three gases as a function of $p$, showing the degree of linearity. The measured values of ${\cal C}_g$ for all gases are a factor of $\sim 5$ larger than the kinetic theory predictions \cite{Supplemental, Reif}.

The data in Fig. 1a  can be fit accurately \cite{Ekinci}.  For a large plate resonator  (${\rm Kn}_l  \approx 0$), the dissipation in a gas of viscosity $\mu_g$ and density $\rho_g$ can be found as \cite{Yakhot_Colosqui, Supplemental}
\begin{equation}\label{eq:plate_soln}
\frac{1}{Q_g}  = \frac{S_r}{m_r}f(\omega_0 \tau )\sqrt {\frac{{\mu_g \rho_g }}{{2\omega_0 }}}.
\end{equation}
Here, $S_r$ is the surface area and $m_r$ is the   mass of the plate resonator, and $f$ is the scaling function \cite{Yakhot_Colosqui} found as $f(x) = \frac{1}{{(1 + x^2 )^{3/4} }}\left[ (1 + x )\cos \left( {\frac{{\tan ^{ - 1} x }}{2}} \right) \right.  \left. - (1 - x)\sin \left( {\frac{{\tan ^{ - 1} x}}{2}} \right) \right]$. The fit in Fig. 1a was obtained using the empirical relation  $\tau = \frac {610 \rm \times 10^{-9}[s \cdot Torr]} {p}$ and experimental parameters \cite{Supplemental}.
%

Now, we turn to   the ``low-frequency limit" of $\omega_0 \tau \to 0$. Figure 2a shows the pressure-dependent dissipation of a low-frequency microcantilever with linear dimensions $l_x \times l_y \times l_z \approx 32 \times 350 \times 1~\mu \rm m^3$  (inset Fig. 2a) and frequency $f_0=18.8$ kHz. We define ${\rm Kn}_l= \frac{\lambda}{l_x}$, as suggested in \cite{Seo, Mertens, Martin}. The transition in Fig. 2a takes place around $p_c \approx 1.2$ Torr, where ${\rm Kn}_l \approx 1$ and $ \omega_0\tau \approx 0.06$. (${\rm Kn}_l \approx 1$ indicates deviation from the low-$p$ molecular asymptote.) The features in Fig. 2a are very similar to those in Fig. 1a: two asymptotes with a well-defined $p_c$. Inspection of the ranges of $\rm Wi$ and ${\rm Kn}_l$ suggests that the transition cannot be tied to frequency ($\rm Wi$) but must be due to the length scale (${\rm Kn}_l$). In other words, the transition from molecular flow (${\rm Kn}_l \gg 1$) to viscous flow (${\rm Kn}_l \ll 1$)   appears to take place around ${\tilde {\rm Kn}_l} = \frac {\lambda_c}{l_x} \approx 1$. While the data trace in Fig. 2a  looks similar to that in Fig. 1a, the transitions observed in the two are due to different physical mechanisms.

In Fig. 2b, we plot the consistently-found $p_c$ in N$_2$ for different sets of devices. Here, the relevant linear dimension $l_x$ is kept constant for each set, but the frequency is varied: diamond nanocantilevers \cite{Kara} with $l_x \approx 800$ nm and $0.4~{\rm MHz} \le f_0 \le 40~{\rm MHz}$;  silicon microcantilevers with $l_x \approx 32~\mu\rm m$ and $14~{\rm kHz} \le f_0 \le 2.4~{\rm MHz}$; and quartz crystals with $l_x\sim 5$ mm and $5~{\rm MHz} \le f_0 \le 75~{\rm MHz}$.  Surprisingly, the linear trend between $p_c$ and $f_0$ holds only for high frequencies, with a saturation at low frequencies. The saturation value of $p_c$ is determined by the condition that $ \lambda \sim \l_x $ (dotted horizontal lines).   The oscillation  frequency (and $\rm Wi$)  becomes the relevant scaling parameter above a certain frequency; at low  frequency, the  length scale  (${\rm Kn}_l$) takes over. Thus, the physics is determined by an interplay between the  relevant length scale  of the body and its  oscillation frequency.

To gain more insight into the transition, we scrutinize ${\tilde {\rm Kn}}_l = \frac {\lambda_c} {l_x}$ and  $\tilde {\rm Wi} = \omega_0 \tau_c$ for each  device at its  $p_c$.  Figure 2c shows  ${\tilde {\rm Kn}}_l$ and  $\tilde {\rm Wi}$ plotted in the $xy$-plane using  logarithmic and linear axes (inset); the dashed lines are  $ \tilde {\rm Wi} + {\tilde {\rm Kn}}_l = 1$. The data suggest that the dissipation is a function of both ${\rm Wi}$ and ${\rm Kn}_l$, and it approximately depends on  $ {\rm Wi}+{\rm Kn}_l$.

\begin{figure}
\centering
\includegraphics[width=3.375in]{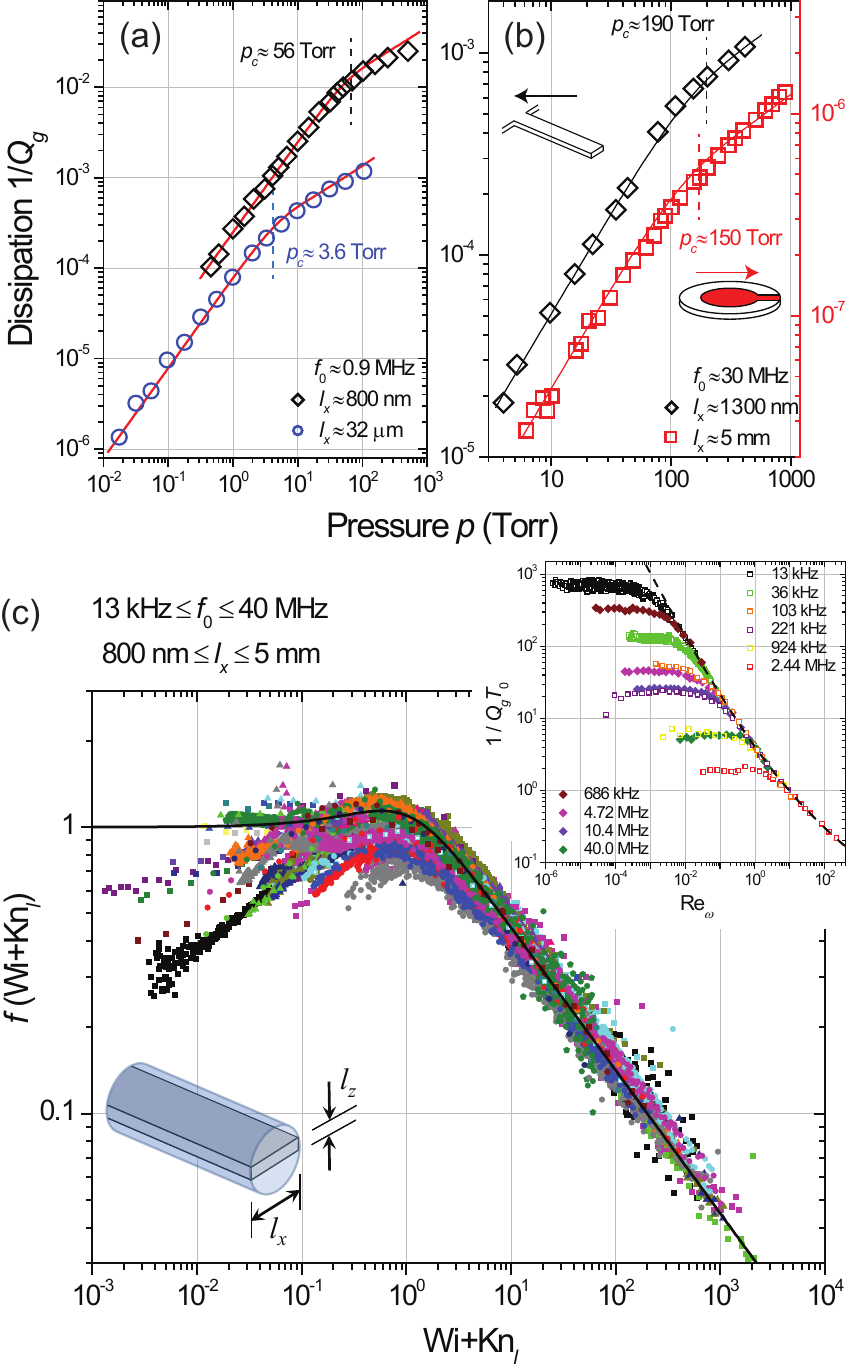}
\caption{(a) Dissipation  \emph{vs.} $p$ for two cantilevers with different length scales but similar frequencies ($l_x \approx 800$ nm, $f_0\approx 894$ kHz; and $l_x \approx 32~\mu$m, $f_0 \approx 924$ kHz) in N$_2$. Transitions are determined by ${\rm Kn}_l \approx 1$  at    $p_c \approx 56$ Torr and 3.6 Torr, respectively. (b) Dissipation for a nanocantilever ($l_x \approx 1300$ nm, $f_0 \approx 28.6$ MHz) and a macroscopic quartz crystal ($l_x \sim 5 $ mm, $f_0 \approx 32.7$ MHz \cite{Ekinci}); the transitions take place around 190 Torr and 150 Torr, respectively. (c) Collapse plot for all the data in different gases. The thick solid line shows the scaling function $f$. The inset is a collapse of select cantilever data based on the viscous cylinder solution. Squares and diamonds correspond to  microcantilevers ($l_x \approx 32~\mu$m) and  nanocantilevers  ($l_x \approx 800$ nm), respectively. Dashed line shows the imaginary part of the complex hydrodynamic function for a cylinder. The lower inset shows the  parameters of the model. }
\label{fig:Qgas}
\end{figure}

We now justify the observed scaling more rigorously by inspecting the stress tensor $\sigma_{ij}$ obtained from the Chapman-Enskog expansion of the Boltzmann equation in the relaxation time approximation. To second order of smallness, the expansion is \cite{Chen}
\begin{equation}\label{eq:2nd_order}
\begin{split}
{\sigma _{ij}} \approx {\sigma _{ij}^{(1)}} + {\sigma _{ij}^{(2)}}  =  & 2\rho_g \theta \Bigl[ \tau {S_{ij}} -  \tau \left( {{{\partial  \over {\partial t}}} + {\bf{u}}\cdot\nabla } \right)(\tau {S_{ij}})  \\  &+ 2 {\tau ^2} \left( {{S_{ik}}{S_{kj}} - \frac{{{\delta _{ij}}}}{3}{S_{kl}}{S_{kl}}} \right) \\ & - 2 {\tau ^2} \left( {{S_{ik}}{\Omega _{kj}} + {S_{jk}}{\Omega _{ki}}} \right) \Bigr].
\end{split}
\end{equation}
As usual, ${S_{ij}} = \frac{1} {2}\left( {\frac{{\partial {u_i}}} {{\partial {x_j}}} + \frac{{\partial {u_j}}}{{\partial {x_i}}}} \right)$ and ${\Omega _{ij}} = \frac{1} {2}\left( {\frac{{\partial {u_i}}} {{\partial {x_j}}} - \frac{{\partial {u_j}}}{{\partial {x_i}}}} \right)$ are the  strain rate and  the vorticity tensors, respectively, with $i,j=x,y,z$; and $\theta = \frac {k_B T} {m_g}$. The last two terms of $\sigma_{ij}$ are the second rank tensor $\xi_{ij}^{(2)}$ of order $(\tau {\bf{S}})^2$, where ${\bf{S}}$ represents the strain rate tensor. There are two dimensionless groups in Eq.~(\ref{eq:2nd_order}): the total time derivative $\tau \frac {d}{dt}=\tau \left(\frac {\partial} {\partial t} + {\bf{u}} \cdot \nabla \right)$ and $\tau {\bf{S}}$. One notices that these two dimensionless groups both remain invariant under Galilean transformations \cite{Supplemental}. In order to satisfy Galilean invariance, therefore, the Chapman-Enskog expansion of kinetic equations must be in powers of these parameters only;  powers of non-Galilean-invariant parameters, e.g., ``bare" $\frac {\partial} {\partial t}$, are forbidden in a flow in an arbitrary geometry.  Accordingly,  one can formally write the Galilean-invariant stress tensor up to all orders as
\begin{equation}\label{eq:nth_order}
\begin{split}
& {\sigma _{ij}} = 2{\rho _g}\theta [\tau {S_{ij}} + \\ & \sum\limits_{n = 2}^\infty  {\left( {{\alpha _{n - 1}}{{( - \tau )}^{n - 1}}{{\left( {{\partial  \over {\partial t}} + {\bf{u}}\cdot\nabla } \right)}^{n - 1}}(\tau {S_{ij}}) + \xi _{ij}^{(n)}} \right)} ].
\end{split}
\end{equation}
Here, $\alpha _{n - 1}$ are constants, and the tensors $\xi _{ij}^{(n)}\sim (\tau {\bf{S}})^n$  are not necessarily zero \cite{expansion_comment}.

A closed form formula can be obtained for the dissipation of a finite-sized body oscillating in a fluid, if the deviations from the infinite plate solution \cite{Yakhot_Colosqui} are assumed  small. As in the infinite plate \cite{Yakhot_Colosqui,Supplemental}, we set all $\alpha_k \approx 1$ and all  $\xi _{ij}^{(n)} \approx 0 $  in Eq.~(\ref{eq:nth_order}).  After non-dimensionalization with ${\bf{\hat u}} = {{\bf{u}} \over c}$, $\hat t = {\omega _0}t$ and $\hat \nabla  = {{l} \nabla}$, the stress tensor ${\sigma _{ij}}$ for a finite-sized body becomes an expansion in powers of the operator $\tau \frac {d} {dt} =  {{\omega _0}\tau {\partial  \over {\partial \hat t}} + {\rm{K}}{{\rm{n}}_l}{\bf{\hat u}}\cdot\hat \nabla }$. The scaling parameter therefore becomes approximately ${\omega_0 \tau +  \rm Kn}_l$, and  the infinite plate solution in Eq.~(\ref{eq:plate_soln}) can be generalized by replacing  $\omega_0 \tau$ with $\omega_0 \tau + {\rm Kn}_l$. Thus, we deduce \cite{Supplemental}
\begin{equation}\label{eq:soln_finite_plate}
\frac{1}{Q_g}  \approx \frac{S_r}{m_r} f(\omega_0 \tau +\frac {\lambda} {l_x})\sqrt {\frac{{\mu_g \rho_g \tau}}{{2(\omega_0 \tau + \frac {\lambda} {l_x})}}}.
\end{equation}
for a finite-sized body oscillating in a fluid. Several points are noteworthy. First, Eq.~(\ref{eq:soln_finite_plate}) is valid in  the asymptotic  and the intermediate ranges. Second, the non-dimensionalization above is eminently reasonable, because the only velocity scale in kinetic theory is the thermal velocity $\sim c$.  Regardless, the dimensional solution is obtained only after imposing the boundary conditions. Finally,  Galilean invariance dictates the form of $\frac{d}{dt}$ and leads to a scaling parameter $\approx {\rm Wi +  Kn}_l$, instead of a more involved combination of ${\rm Wi}$ and ${\rm Kn}_l$.

A number of fits to experimental data using Eq.~(\ref{eq:soln_finite_plate}) are shown in Figs. 2a, 3a, and 3b as well as in the Supplemental Material \cite{Supplemental}. The data in Fig. 3a and 3b are examples of the low- and high-frequency limits, respectively. Here, different-sized but similar-frequency resonators are compared. All fits are obtained as follows. First, $S_r/m_r$ is determined from linear dimensions or from separate measurements when necessary \cite{Supplemental}. For each pressure, the value of $\omega_0 \tau + \frac{\lambda}{l_x}$ is computed using  $\tau={\cal C}_g /p$ and  $\lambda \approx 0.23\frac{k_{B}T}{{d_g}^{2}p}$ of the gas, and $\l_x$ and   $\omega_0$ of the resonator. Finally, the dissipation is found from Eq.~(\ref{eq:soln_finite_plate}) at each pressure using tabulated  $\mu_g$ and $\rho_g$,  and our empirical $\tau$. To improve the fits, the theoretical prediction is multiplied by an ${\cal O} (1)$ constant ${\cal Q}_p$ .  The collapse plot in Fig. 3c is obtained by properly dividing the data by $\frac{S_r}{m_r} \sqrt {\frac{{\mu_g \rho_g \tau}}{{2(\omega_0 \tau+ \frac {\lambda} {l_x})}}}{\cal Q}_p $ and  plotting the results as a function of  $\omega_0\tau+\frac{\lambda}{l_x}$.  The thick solid line shows $f({\rm Wi + Kn}_l)$.  There are no free parameters other than the  fitting factors ${\cal Q}_p$ with mean $\bar{\cal Q}_p \approx 2.6 \pm 0.5$ \cite{Supplemental}.

At the viscous limit ${\rm Wi + Kn}_l \ll 1$, the cantilever data deviate from the plate solution and converge to a cylinder solution. The cylinder solution yields $\frac{1}{{{Q_g}}} \approx \frac{{{\Gamma _I}({\rm Re}_{\omega} )}} {1/T_0+{{\Gamma _R}({\rm Re}_{\omega} )}}$ \cite{Sader, Paul}. Here,  $\Gamma ({\rm Re}_{\omega})= \Gamma_R ({\rm Re}_{\omega})+i \Gamma_I ({\rm Re}_{\omega}) $  is the complex hydrodynamic function for a cylinder and only depends upon the (oscillatory) Reynolds number ${\rm Re}_{\omega} = \frac {\omega_0 {l_x}^2} {4 \nu_g}$; $T_0=\frac {\pi}{4} \frac{\rho_g l_x}{\rho_r l_z}$ with $\rho_r$ being the density of the solid (Fig. 3c lower inset). For our gas experiments, $1/T_0 \gtrsim 1000 \gg \Gamma_R$, and  thus  $\frac{1}{{{ Q_g T_0}}} \approx \Gamma _I({\rm Re}_{\omega})$. The upper inset of Fig. 3c shows $\frac{1}{{{ Q_g T_0}}}$ from representative cantilevers with different parameters plotted against ${\rm Re}_{\omega}$;  dashed line shows $\Gamma_I({\rm Re}_{\omega})$. In each case, a fitting constant ${\cal Q}_c$ with mean ${\bar{\cal Q}_c} \approx 0.9 \pm 0.2$ is used \cite{Supplemental}. The  data converge to the cylinder solution in the viscous regime.

We conclude that the scaling parameter for an arbitrary time-dependent isothermal flow should be a function of both $\rm Wi$ and ${\rm Kn}_l$. We show   that a generalized Knudsen number in the form ${\rm Wi+Kn}_l$ works well and can be justified by Galilean invariance.


\begin{acknowledgments}
We acknowledge partial support from US NSF (through Grant No.  CBET-1604075).
\end{acknowledgments}



\begin{thebibliography}{99}
\bibitem{expansion_comment} Depending on the flow problem, the expansion may contain other terms, such as mixed powers of time and space derivatives. Such terms are omitted here for clarity.
\end{thebibliography}
\end{document}


\title{Supplementary Material for\\ ``Generalized Knudsen Number for Unsteady Fluid Flow''}

\author{V. Kara}
%

\author{V. Yakhot}

\author{K. L. Ekinci}
\affiliation{\BU}

\date{\today}
\maketitle

\section{Properties of the Resonators}
In our experiments, we used different resonators with different dimensions and frequencies. For a few of these resonators, we employed both fundamental and harmonic modes. Tables  \ref{micro}, \ref{nano}, \ref{quartz}, and \ref{membrane} summarize the parameters of these devices along with their transition pressures $p_c$ in different gases. The first set of devices are silicon microcantilever resonators, which are listed in Table \ref{micro}; these are tipless microcantilevers from Mikromasch (NSC36, CSC38, and NSC15 series).  All of the microcantilevers have widths $l_x$  around $30-35~\mu\rm m$, with the exception of two that were  obtained by focused ion beam (FIB) milling.  Table \ref{nano} shows the  parameters for  our single-crystal diamond  nanocantilevers. The fabrication  process for these nanocantilevers as well as fluidic data for a  select few were published previously \cite{Kara}. These nanocantilevers  have characteristic triangular cross-sections; here, $l_x$  is the width and  $l_z$ is the height of the triangular cross-section; there is also a 6-$\mu$m gap between the nanocantilevers and the substrate \cite{Kara}. Because the gaps  are large here,  squeeze damping becomes mostly irrelevant \cite{Lissandrello, Bhiladvala_Knudsen, Bao_Squeeze}. Table \ref{quartz} shows the  parameters for our quartz crystal resonators. The diameters  of the metal electrodes on the crystals  (see inset of Fig. 1a in the main text) \emph{roughly} correspond to the size of the flow \cite{Herrscher}.  We also measured a few modes of a thin square silicon nitride membrane (Table \ref{membrane}), which was tension dominated.

\begin{table}
\caption{Some properties of the silicon microcantilevers used in this work. The second column is the mode; the third column shows the resonance frequency of the mode; the linear dimensions are listed in the fourth column (see inset of Fig. 2a in main text);  $p_c$ is the transition pressure. FIB milled microcantilevers are marked with *;  silane coated microcantilevers are marked with $\dagger$. }
\centering
\begin{tabular*}{\textwidth}{c @{\extracolsep{\fill}} cccccc}
\hline
& \multicolumn{6}{c}{Silicon Microcantilevers } \\
\hline
Device	& 	Mode	& 	$f_{0}$	&$l_x \times l_y \times l_z$ & 	 &$p_{c}$ &\\
	& 		& 	 (kHz)	& $ (\mu \rm m^3)$& 	 & (Torr)&\\
\hline
& & & &N$_{2}$ &He&Ar\\
\hline
\midrule
C1&	1&13.7&$40  \times  230  \times  3$	& $1.1$ 	& $-$		&$-$\\
C2&1	&14.2	&$32  \times  350  \times  1$	& $0.9$ 	& $2.1$		&$1.1$\\
C3$^*$ & 1 & 17.3& $14\times350\times1$ 			&$2.0$		&$5.2$	& -\\
C4$^\dagger$ & 1 & 18.8 & $32\times350\times1$ &$1.2$   &$2.7$  & $1.3$\\
C5$^\dagger$ & 1 & 25.6 & $32\times350\times1$ &$1.2$&$2.5$&$1.1$\\
C6 & 1 & 27.9 & $32  \times  250  \times  1$  &$1.0$ & $2.7$ & $1.5$\\
C7$^\dagger$ & 1 & 36.4 & $32\times250\times1$ &$1.7$&$2.8$&$1.2$\\
C8& 1 & 78 & $53\times460\times2$ &$1.4$&-&-\\
C9 & 1 & 103& $32  \times  130  \times  1$  &$1.5$ & $3.8$ &$1.9$\\
C9 & 2 & 638& $32  \times  130  \times  1$  &$2.7$ &$6.4$& $3.3$\\
C9& 3 & 1762& $32  \times  130  \times  1$  &$4.6$&$10$&$9.5$\\
C10 & 2 & 676& $32  \times  130  \times  1$  	&$2.5$ & - &  - \\
C10& 3 & 1864& $32  \times  130  \times  1$  	&$5.0$& - &  - \\
C11 & 1 & 143& $32  \times  110  \times  1$  &$3.0$&$5.0$&$1.7$\\
C11 & 3 & 2438& $32  \times  110 \times  1$  &$10$&$15$& $13$\\
C12 & 1 & 150& $32  \times  110  \times  1$  	&$2.2$ & - & - \\
C12 & 2 & 924& $32  \times  110  \times  1$  	&$3.6$  & - & - \\
C13 & 1 & 216& $32  \times  90  \times  1$  & - & $5.6$	& $2.4$\\
C13 & 2 & 1311&  $32  \times  90  \times  1$  & - & $12$	&$5.0$\\
C14 & 1 & 221& $35\times110\times2$ 			&$1.9$ & - & -\\
C14 & 2 & 1360& $35\times110\times2$ 		&$4.6$ & - & -\\
C15 & 1 & 223& $32  \times  90  \times  1$  	&$2.2$ & - & -\\
C15 & 2 & 1356& $32  \times  90  \times  1$  	&$4.9$& - &  - \\
C16& 1 & 310& $36\times125\times3.6$& 2.1  &$-$& - \\
C17& 1 & 360& $30\times125\times4$& -  &$4.9$& - \\
C18$^*$ & 1 & 5780& $8\times30\times4$ &$35$ &$48$& $37$\\
\hline
\label{micro}
\end{tabular*}
\end{table}

\begin{table}
\caption{Single-crystal diamond nanocantilevers. These have triangular cross-sections; additionally, there is a gap of $ 6~\mu$m between the cantilever and the substrate. More details on these structures can be found in \cite{Kara}. }
\centering
\begin{tabular*}{\textwidth}{c @{\extracolsep{\fill}} cccccc}

\hline
& \multicolumn{6}{c}{Single-crystal Diamond Nanocantilevers} \\
\hline
Device	& 	Mode	& 	$f_{0}$ 	& $l_x \times l_y \times l_z$  & &$p_{c}$ &\\
	& 		& 	 (MHz)	&  $(\mu \rm m^3)$ & & (Torr)&\\
\hline
& & & &N$_{2}$ &He&Ar\\
\hline

D1 & 1 & 0.411& $0.820   \times 48  \times  0.530$	&$44$	&$130$	& $37$\\
D2 & 1 & 0.539& $0.820 \times43  \times  0.530$ 	&$46$	&-		&-\\
D3 & 1 & 0.686 & $0.820 \times38 \times  0.530$ 	&$47$ 	&-		& -\\
D4 & 1 & 0.894 & $0.820 \times34  \times  0.530$ 	&$56$ 	&-		&-\\
D5 & 1 & 1.211 & $0.820   \times  29 \times  0.530$  &$44$ 	&$130$	& $52$\\
D6 & 1 & 1.735 & $0.820 \times24 \times  0.530$ 	&$50$ 	&-		&-\\
D7 & 1 & 2.691& $0.820   \times 19 \times  0.530$ 	&$50$ 	&$135$	&-\\
D8 & 1 & 4.725& $0.820   \times  14.7  \times  0.530$	&$52$ 	&-		&-\\
D9 & 1 & 10.421& $0.820   \times 9.6  \times  0.530$ &$64$ 	&$165$	&$76$\\
D10 & 1 & 28.653& $1.300  \times  9.6  \times  1.150$ &$187$ &-	& -\\
D11 & 1 & 40.032& $ 0.820  \times  4.8  \times  0.530$  &$211$&$448$	& $240$\\
\hline
\label{nano}
\end{tabular*}
\end{table}
\begin{table}
\caption{ Properties of the quartz crystals. As described in the main text, $l_x$  stands for the diameter of the metal electrode on the crystal and  $l_z$ for the thickness of the crystal. For three of these devices, $S_r/m_r$ values were directly obtained from experiments (see below and \cite{EkinciPRL}).}
\centering
\begin{tabular*}{\textwidth}{c @{\extracolsep{\fill}} ccccccc}
\hline
& \multicolumn{6}{c}{Quartz Crystals } \\
\hline
Device& Mode& $f_{0}$  & $l_x \times l_z $ & $S_r/m_r $ & &$p_{c}$&\\
& &  (MHz) &  $(\rm mm^{2})$ & $ \rm ( m^2/kg)$ & &(Torr)&\\
\hline
& & & & &N$_{2} $&He&Ar\\
\hline
Q1& 1 & 5 & $5 \times 0.12$	&2.7 &$18$ &$26$& $27$\\
Q1& 2 & 15&$ 5\times 0.12$	&- &$88$ &$96$ &$86$\\
Q2& 1 & 6 & $9\times 0.12$&	-	&$22$ &- &-\\
Q2& 2 & 18& $9\times 0.12$	& -	&$78$  &$46$ &-\\
Q3& 1 & 14.3&$ 10\times 0.1$& 9.8 &$70$  &-& -\\
Q4& 1 & 25&$ 6.5\times 0.1$	& - &$101$&$84$& $133$\\
Q4& 2 & 75&$ 6.5\times 0.1$	& - &$258$&$256$& $287$\\
Q5& 1 & 32.7&$ 5\times 0.1$& 6.3 &$146$ &-& -\\
Q6& 1 & 50&-& - &$158$&-&-\\
\hline 
\label{quartz}
\end{tabular*}
\end{table}
\begin{table}
\caption{Properties of the silicon nitride membrane. This is a square membrane. More details can be found in \cite{SukumarPRL}. }
\centering
\begin{tabular*}{\textwidth}{c @{\extracolsep{\fill}} cccccc}
\hline
 &\multicolumn{5}{c}{Silicon Nitride Membrane} \\
 \hline
\cmidrule(l){2-6}
Device& Mode& $f_{0}$ &$l_x \times l_y\times l_z $  & &$p_{c}$&\\
& &  (kHz)&  $ (\rm \mu m^3)$ & &(Torr)&\\
\hline
& & & &N$_{2} $&He&Ar\\
  \hline
\midrule
M1 & (6, 6) & 885& $1260  \times  1260\times0.53$ &$6.8$&-& -\\
M1 & (9, 9) & 1330& $1260  \times  1260\times0.53$ &$8.6$&-& -\\
M1 & (10, 10) & 1479& $1260  \times  1260\times0.53$ &$6.3$&-&-\\
\hline
\label{membrane}
\end{tabular*}
\end{table}

The typical amplitudes of oscillations remained deep in the linear regime for all the cantilevers, as determined from optical interferometry. For the low-frequency microcantilevers, the maximum amplitude was $\sim 10$ nm. Most single-crystal diamond nanocantilevers were driven at an amplitude  $\lesssim 0.1$ nm. In a few experiments, the dissipation was extracted from thermal motion, in which the r.m.s amplitude remained $\lesssim 0.01$ nm. The amplitudes of the quartz resonators were estimated to be in the range $1-10$ nm.

When gas molecules collide with a solid surface, they are reflected either diffusively or specularly  from the surface. Different surface properties and/or surface adsorbates (e.g., water) may lead to different surface accommodation and hence dissipation. Surface accommodation becomes more relevant at low pressures when the mean free path of the gas is large \cite{Ducker_2010}. In order to understand the effect of surface properties on the dissipation, we coated some of the microcantilevers (C4, C5, and C7) with silane, which made the surface hydrophobic and reduced the amount of adsorbed water and other contaminants \cite{honig-ducker}. While the absolute value of the dissipation was expected to depend on the surface accommodation, we did not observe any measurable differences in the \emph{functional form} of the dissipation and the transition pressure $p_c$, to within our error. In a separate study, to get rid of the adsorbed water, we mildly baked the 5-MHz quartz crystal resonator in vacuum ($10^{-5}$ Torr)  at a temperature of $\sim 150^{\circ}$C for 2 hours.  We performed experiments with the fundamental and first harmonic of this resonator in N$_2$ and He, but  again did not observe any significant differences between the measurements on baked and unbaked resonators.

\section{ Measurements}

\subsection{Vacuum chamber}

A  vacuum system  with optical access is used in most experiments. The samples (cantilever and membrane resonators) are placed inside the vacuum chamber after they are glued on top of a piezo-actuator.  The pressure of the  chamber is monitored with two capacitive gauges (MKS Baratron $722$A) between $10^{-3}-10^3$ Torr continuously; the response of these gauges are independent of the gas used. There are several other gauges attached to the system for monitoring lower pressures.  The vacuum system is pumped by a turbo pump that brings the pressure from atmosphere down to $p\approx 10^{-5}$ Torr. There is also  an ion pump which pumps the chamber  down to $p\approx 10^{-7}$ Torr. When we start the pressure sweeps, we close  all the valves and admit gas into the chamber while monitoring the pressure. The gas lines are pumped and flushed in order to limit contamination.

We measure the intrinsic quality factor $Q_{0}$ at the lowest pressure, typically $p\approx 10^{-7} - 10^{-6}$ Torr. For this measurement, we use a Network Analyzer (NA) or a Spectrum Analyzer (SA) to detect the driven response or the thermal fluctuations of the resonator, respectively.  In our electrical measurements of quartz crystals, we use a second bakeable vacuum chamber that is connected to the same pumping system.

\subsection{ Optical Measurements}

We use a homodyne optical interferometer for detecting the response of the cantilevers and membrane resonators. A stabilized HeNe laser with 1 mW peak power  is operated in its intensity stabilization mode. One photodetector (Thorlabs PDA8A) is used for path stabilization.  A second photodetector (New Focus 1801) is used  to detect the oscillations of the resonator. In a typical experiment, $\lesssim 100~\mu$W of power is incident on the resonator. With $\sim 40~\mu$W of power on the photodetector, we obtain a displacement sensitivity of $\sim 30$ fm$/\sqrt{\rm Hz}$. We do not notice any sample heating from the detection laser.

To measure the gas dissipation, we use two different approaches. In the first, we directly measure the frequency response of the resonator at different $p$.  We then fit  the  line-shapes to Lorentzians in order to obtain the dissipation and the resonance frequencies as a function of $p$. Alternatively,  we use a standard phase locked loop (PLL) to track the amplitude of the oscillations at resonance under constant drive as a function of $p$. The measured amplitude is assumed to be proportional to the quality factor. We then convert the amplitude data to gas dissipation using quality factors measured at the lowest and highest pressures.

\begin{figure*}
\centering
\includegraphics[width=6.75in]{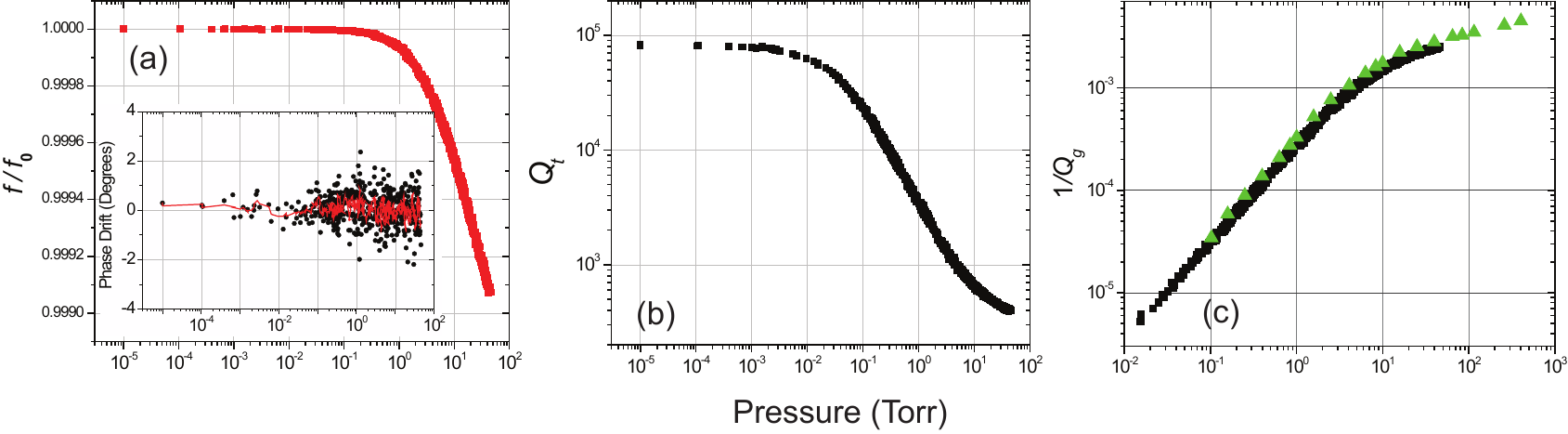}
\caption{A typical set of experimental results obtained on C9 in He gas. (a) The resonance frequency shift as a function of $p$ obtained using a PLL measurement. The inset shows the deviations from the locked phase during the pressure sweep. (The red curve is the smoothed signal.)  (b) The total quality factor $Q_t$ as a function of $p$. As the quality factor degrades at high pressures, the phase fluctuations shown in (a) increase.  (c) The gas dissipation as a function of $p$ obtained from the data in (b). The black squares are the results from the PLL measurement. The green triangles are obtained from NA sweeps. At each pressure, the resonator response is obtained. This response is fit to a Lorentzian, which provides the quality factor (after proper intrinsic dissipation subtraction). }
\label{PLL}
\end{figure*}

Figure \ref{PLL}  shows  typical experimental results. Figure \ref{PLL}a is the  resonance frequency shift  of C9 as a function of pressure, measured using the PLL technique. The maximum shift  is around 0.1\%  here. (In all the experiments, the frequency shift remained less than  0.3\%.) This suggests that the mass loading  due to the fluid  is negligible. The inset in  Fig. \ref{PLL}b  shows the phase fluctuations in the PLL during the sweep. Figure \ref{PLL}b shows the total quality factor $Q_t$ in the same experiment.  At very low pressures, the quality factor becomes independent of the pressure and is dominated by intrinsic loss mechanisms, converging to $Q_0$. Having obtained  $Q_0$, we extract the pressure-dependent dissipation  using the relation, $\frac{1}{Q_g}=\frac{1}{Q_t}-\frac{1}{Q_0}$. The  outcome of this step is shown in Figure \ref{PLL}c. Here, two data sets, one obtained using the PLL technique and the other using network analysis, are compared.

\subsection{Electrical Measurements}

To measure the pressure-dependent dissipation of quartz crystal resonators, we monitor the (mechanical) impedance of the resonator \cite{Johannsmann}  around its mechanical resonance. In the measurement, a constant drive voltage is applied on the resonator  as the pressure is varied; the drive frequency (phase) is always kept around the  resonance using a  PLL. The measured mechanical resistance is linearly proportional to the dissipation \cite{kokubun}. For some devices, we converted the electrical signal to quality factor using a ring-down measurement \cite{Borovsky}. In the ring-down measurement, the resonator was driven by a sinusoidal voltage; the voltage was turned off using an electrical switch; and the ring-down signal was detected using a digital oscilloscope. The effects of the input impedance of the measurement circuit were taken into account. Fitting the ring-down signal to an exponential decay provided the quality factor. For  devices, for which the quality factor was not measured or the $S_r/m_r$ was not available,  the dissipation results are left in arbitrary units. Thus, fits to these data sets are multiplied by arbitrary constants. Regardless, $p_c$ values are still accurate.


\section{Theory and Fits}

\subsection{Kinetic Theory}

Kinetic theory is used to predict the mean free path $\lambda$ of the gases used in this work.  We use $\lambda \approx \frac{{{k_B}T}}{{\sqrt 2 \pi {{d_g}^2}p}}$, where $k_BT$ is thermal energy, $d_g$ is the diameter of a gas molecule, and $p$ is the pressure. In the calculations, the molecular diameters for He, N$_2$ and Ar are taken as 220, 375, and 364 pm, respectively \cite{hanlon}. Similarly, the relaxation time $\tau$ can be estimated from kinetic theory as $\tau \approx \frac {\lambda}{\bar{v}_{th}} $ where $\bar{v}_{th} = \sqrt{\frac{8k_B T}{\pi m_g}}$ with $m_g$ being the mass of a gas molecule. Using the expression for $\lambda$, the relaxation time can be written as a function of pressure as $\tau = \frac {{\cal C}_g}{p}$, where ${\cal C}_g =\frac {\sqrt{m_g k_B T}}{4\sqrt{\pi} {d_g}^2}$ is a gas dependent constant. With the molecular diameters  above, kinetic theory yields ${\cal C}_{\rm He}=  113\times 10^{-9}$, ${\cal C}_{\rm N_2}= 103 \times 10^{-9}$,  and ${\cal C}_{\rm Ar}= 131 \times 10^{-9}$, all in units of s$\cdot$Torr. The respective empirical values (found as described in the main text) are: ${\cal C}_{\rm He}= 560 \pm 70 \times 10^{-9}$, ${\cal C}_{\rm N_2}= 610 \pm 30 \times 10^{-9}$,  and ${\cal C}_{\rm Ar}= 750 \pm 80 \times 10^{-9}$, all in units of s$\cdot$Torr. We note  that the empirical relationship for $\tau$ that we use here is slightly different from that in \cite{EkinciPRL}. The reason is the arbitrariness in the determination of the transition pressure $p_c$. Here, we used a fixed deviation from the low-$p$ (molecular) asymptote to indicate $\omega_0 \tau \approx 1$.

As mentioned in the main text, the measured values of ${\cal C}_g$ for all gases are a factor of $\sim 5$ larger than the kinetic theory predictions. This may be due to the fact that it takes a few collisions  to establish equilibrium  or that relaxation near a wall takes  longer than that in bulk. These findings suggest that mean free path $\lambda$ should also deviate from the kinetic theory prediction near the wall. A direct measurement of $\lambda$ near the wall could provide interesting insights into this question. In summary, the near-wall dynamics of gases appears to deviate from kinetic theory, and more theoretical and experimental studies are needed to resolve these subtle deviations.

\subsection{Galilean Invariance of the Total Time Derivative}

Equations of Newtonian mechanics must remain invariant when written down by  observers in different inertial reference frames. Here, we  review this principle of Galilean invariance and show that $\frac {d}{dt}$ remains invariant under a Galilean transformation. For simplicity, we  consider a one dimensional problem, in which two inertial frames are related by a Galilean transformation: $x=x'+Ut'$ and $t=t'$. Here, $U$ is the (constant) velocity of the primed frame with respect to the unprimed one. We are concerned with a (scalar) field $\chi$, which is a function of both position and time. In order to relate the derivatives, we consider a change in $\chi$ as
\begin{equation}
d\chi = {{\partial \chi} \over {\partial t}}dt + {{\partial \chi} \over {\partial x}}dx = {{\partial \chi} \over {\partial t'}}dt' + {{\partial \chi} \over {\partial x'}}dx'.
\end{equation}
We notice that $dt'=dt$, and $dx'=dx-Udt$. Substituting the differentials in $d\chi$, we   find that ${\partial  \over {\partial x}} = {\partial  \over {\partial x'}}$ and ${\partial  \over {\partial t}} = {\partial  \over {\partial t'}} - U{\partial  \over {\partial x'}}$.

For a flow,   the total time derivative (also called substantial derivative) is given by ${d \over {dt}}={\partial  \over {\partial t}} + u{\partial  \over {\partial x}}$,  where $u$ is the  velocity field of the flow (as measured in the unprimed frame). Galilean transformation of this operator to the primed frame results in
\begin{equation}
{\partial  \over {\partial t}} + u{\partial  \over {\partial x}} = {\partial  \over {\partial t'}} + (u - U){\partial  \over {\partial x'}},
\end{equation}
where $u-U=u'$ is now the  velocity field measured in the primed frame. It can be seen that
\begin{equation}
{\partial  \over {\partial t}} + u{\partial  \over {\partial x}} = {\partial  \over {\partial t'}} + u'{\partial  \over {\partial x'}},
\end{equation}
and the total time derivative ${d \over {dt}}$  remains invariant under a Galilean transformation, as expected.  To obtain a Galilean invariant ${d \over {dt}}$,  one therefore needs the presence of both the partial time derivative  and the convective derivative. This, in turn, leads to the scaling parameter observed in this work. (We note that similar steps can be used to show the Galilean invariance of $S_{ij}$.)

\subsection{Plate Theory}

Yakhot and Colosqui \cite{Yakhot_Colosqui} derived the infinite plate result presented in the main text. Here, we first outline their results, and then deduce a closed-form formula for the dissipation of a finite-sized body oscillating in a fluid.

As discussed in the main text, one can formally expand the stress tensor up to all orders as
\begin{equation}\label{eq:stress_general}
{\sigma _{ij}} =  2\rho_g \theta \left[ \tau {S_{ij}} +
 \sum\limits_{n = 2}^\infty  {\left\{ {{\alpha _{n - 1}}{{( - \tau )}^{n - 1}}{{\left( {{{\partial  \over {\partial t}}} + {\bf{u}}\cdot\nabla} \right)}^{n - 1}} (\tau {S_{ij}}) + \xi _{ij}^{(n)}} \right\}}  \right].
\end{equation}
Here, $\alpha _{n - 1}$ are constants, and the tensors $\xi _{ij}^{(n)}\sim (\tau {\bf{S}})^n$  are not necessarily zero. All the other symbols have been defined in the main text. For an infinite plate oscillating in the $x$ direction and with the $z$ coordinate being normal to the plate (see main text and Fig. 1a inset; different from the coordinate system used in \cite{Yakhot_Colosqui}), the only non-zero velocity component $u_x$ of the fluid is along the $x$ direction,  and $ \frac {\partial u_x}{\partial z}$ is the only non-zero spatial velocity derivative. Because of the symmetries discussed in \cite{Yakhot_Colosqui}, all  tensors $\xi_{ij}^{(n)}$ of order $(\tau {\bf{S}})^n$ are zero. Furthermore, ${\bf{u}}\cdot\nabla =0$.  Thus, the  stress tensor simplifies to
\begin{equation}\label{eq:stress_plate}
{\sigma _{xz}} =  2\rho_g \theta \tau \left[  {S_{xz}} +  \sum\limits_{n = 2}^\infty  { {{}{{( - \tau )}^{n-1}}{{\left( {{{\partial  \over {\partial t}}}} \right)}^{n - 1}} {S_{xz}} } }  \right],
\end{equation}
where ${S_{xz}}= \frac{1}{2} \frac {\partial u_x}{\partial z}$. Equivalently, using the dimensionless time $\hat t = {\omega _0}t$,
\begin{equation}\label{eq:stress_plate_dimensionless}
{\sigma _{xz}} = 2{\rho _g}\theta \left[\tau {S_{xz}} + \sum\limits_{n = 2}^\infty  {{{( - 1)}^{n - 1}}{{\left( {{\omega _0}\tau {\partial  \over {\partial \hat t}}} \right)}^{n - 1}}\left( {\tau {S_{xz}}} \right)} \right].
\end{equation}
Again because of the symmetries, the equation to be solved simplifies to
\begin{equation}\label{eq:BGK}
{{\partial u_x} \over {\partial t}} = {1 \over \rho_g }{\partial  \over {\partial z}}{\sigma _{xz}}.
\end{equation}
After combining all the above elements, inverting the operator  containing the infinite series of time derivatives and with $\nu\approx \theta \tau$, one obtains
\begin{equation}\label{eq:telegrapher}
\tau {{{\partial ^2} u_x} \over {\partial {t^2}}} + {{\partial u_x} \over {\partial t}} = \nu {{{\partial ^2}u_x} \over {\partial {z^2}}}.
\end{equation}
The velocity field can be obtained by solving  Eq.~(\ref{eq:telegrapher}) analytically. Using the no-slip boundary condition on the plate, one can subsequently find the energy dissipated by the plate. This is Eq.~(1) of the main text, re-written below as
\begin{equation}\label{eq:plate_soln}
{1 \over {{Q_g}}} = {{{S_r}} \over {{m_r}}}{{f({\omega _0}\tau )} \over {\sqrt {{\omega _0}\tau } }}\sqrt {{{{\mu _g}{\rho _g}\tau } \over 2}}
\end{equation}
with $f$ being the scaling function
\begin{equation}\label{eq:f}
f(x ) = \frac{1}{{(1 + x^2 )^{3/4} }}\left[ (1 + x )\cos \left( {\frac{{\tan ^{ - 1} x }}{2}} \right)  - (1 - x)\sin \left( {\frac{{\tan ^{ - 1} x}}{2}} \right) \right].
\end{equation}
Note that the only dimensionless group in Eq.~(\ref{eq:plate_soln}) is $\omega_0 \tau$, which also appears in the stress expansion in Eq.~(\ref{eq:stress_plate_dimensionless}).  (An alternative derivation for the infinite plate problem is given in \cite{Ekinci_LOC}.)

To find an approximate solution for the dissipation of a finite-sized body, we re-inspect $\sigma_{ij}$ in Eq.~(\ref{eq:stress_general}). The assumption here is that the deviations from the infinite plate are small; thus, all $\alpha_k=1$ and  all  $\xi _{ij}^{(n)} \approx 0$. Using the dimensionless quantities ${\bf{\hat u}} = {{\bf{u}} \over c}$, $\hat t = {\omega _0}t$ and $\hat \nabla  = {{l} \nabla}$, we obtain
\begin{equation}\label{eq:stress_general_approximate}
{\sigma _{ij}} \approx  2\rho_g \theta \left[ \tau {S_{ij}} +
 \sum\limits_{n = 2}^\infty  {\left\{ {{}{{( - 1)}^{n - 1}}{{\left( {{\omega_0 \tau {\partial  \over {\partial {\hat t}}}} + {\rm{K}}{{\rm{n}}_l}{\bf{\hat u}}\cdot\hat \nabla } \right)}^{n - 1}} (\tau {S_{ij}}) } \right\}}  \right],
\end{equation}
which is an expansion in powers of the operator $ {{\omega _0}\tau {\partial  \over {\partial \hat t}} + {\rm{K}}{{\rm{n}}_l}{\bf{\hat u}}\cdot\hat \nabla } $. A comparison between Eq.~(\ref{eq:stress_general_approximate}) and Eq.~(\ref{eq:stress_plate_dimensionless}) suggests that the dimensionless parameter (scaling parameter) in the  finite-sized body problem is $\omega_0 \tau + {\rm Kn}_l$. By direct analogy to the infinite plate solution in Eq.~(\ref{eq:plate_soln}), we come up with the approximate solution
\begin{equation}\label{eq:soln_finite_plate}
{1 \over {{Q_g}}} \approx {{{S_r}} \over {{m_r}}}{{f({\omega _0}\tau  + {\lambda  \over {{l_x}}})} \over {\sqrt {{\omega _0}\tau  + {\lambda  \over {{l_x}}}} }}\sqrt {{{{\mu _g}{\rho _g}\tau } \over 2}}.
\end{equation}
In Eq.~(\ref{eq:soln_finite_plate}) as well as in Eq.~(\ref{eq:plate_soln}), $S_r$,  $m_r$, and $\omega_0=2\pi f_0$ are the surface area, mass and resonance frequency of the resonator, respectively; and $\tau$, $\lambda$, $\mu_g$, and $\rho_g$ are the relaxation time, mean free path, viscosity and density of the gas, respectively.

After we measure the fluidic dissipation for each resonator, we fit the data to our theory. In all the results  shown below in Figs.~\ref{N2silicon}-\ref{Hequartz},  the red solid line shows the theoretical predictions of Eq.~(\ref{eq:soln_finite_plate}). For cantilevers, we calculated $S_r$ and $m_r$ from their linear dimensions. For silicon microcantilevers, which have rectangular cross-sections,  $S_r \approx 2 l_x (l_y + l_z)$ and $m_r \approx \rho_r l_x l_y l_z$, where $\rho_r$ is the density of the solid (see the illustration in the inset of Fig. 2a in the main text for the linear dimensions).  Singe-crystal diamond nanocantilevers have  triangular cross-sections \cite{Kara}, and $S_r \approx l_x l_y+2 l_y \sqrt{\frac{l_x^2}{4}+l_z^2}$ and $m_r \approx \rho_r\frac{l_x l_yl_z}{2}$. It is not straightforward to calculate $S_r$ and $m_r$ for quartz crystals. For some quartz crystal resonators, we deposited a known thickness of gold film on the crystal and measured the resonance frequency shift. Then, using the Sauerbrey formula \cite{vogt}, we  estimated $S_r/m_r$ from the gold thickness and frequency shift \cite{EkinciPRL}. It must also be noted that the thickness of some of these crystals were not uniform, increasing the errors in the estimates. In order to obtain better fits, we multiplied the theory in Eq.~(\ref{eq:soln_finite_plate}) with a constant of order one. This constant ${\cal Q}_p$ is most probably due to geometric effects  because  Eq.~(\ref{eq:soln_finite_plate}) is deduced from the infinite plate solution in Eq.~(\ref{eq:plate_soln}).  Where available, these fitting constants ${\cal Q}_p$ are shown on the plot. For all the fits, the goodness of fit, as determined from an $R^2$ measurement, was routinely $\gtrsim 0.95$. In some cantilevers, the deviation from the plate theory was more significant in the viscous regime and prompted us to consider the cylinder theory.

\subsection{Cylinder Theory}

For the cantilever resonators,   the  plate theory deviates from experimental data at high $p$, when the resonators enter the Newtonian flow regime. Here, geometry becomes important. For these slender structures, a cylinder approximation is typically used in this limit. For a long cylinder oscillating in a fluid, the dissipation is given as \cite{sader, paul}
\begin{equation}\label{sader}
\frac{1}{Q_g} = \frac{\Gamma_I({\rm Re}_\omega)}{\frac{1}{T_0}+\Gamma_R({\rm Re}_\omega)},
\end{equation}
where $\Gamma_I({\rm Re}_\omega)$, and $\Gamma_R({\rm Re}_\omega)$ are respectively the imaginary and the real parts of the hydrodynamic function, $\Gamma({\rm Re}_\omega)$,
\begin{equation}\label{stokes}
\Gamma({\rm Re}_\omega) =1+ \frac{4iK_1(-i\sqrt{i{\rm Re}_\omega})}{\sqrt{i{\rm Re}_\omega}K_0\left(-i\sqrt{i {\rm Re}_\omega}\right)}.
\end{equation}
In the above expressions, ${\rm Re}_\omega=\frac{\rho_g \omega_0 {l_x}^2}{4\mu_g}$ is the oscillatory Reynolds number. The constant $T_0$ compares the mass of the fluid to that of the solid: $T_0=\frac{m_f}{m_c}=\frac{\pi}{4}\frac{l_x}{l_z}\frac{\rho_g}{\rho_c}$ (see the lower inset of Fig. 3c in the main text for linear dimensions). The functions $K_1$ and $K_0$  in Eq.~(\ref{stokes}) are modified Bessel functions of the third kind \cite{sader}.

In  the figures below,  the dark cyan solid curve  shows the cylinder fits to  experimental data in the Newtonian regime obtained using the above expressions. As in the plate case, we multiplied the theory with a constant of order one in order to improve the fits. These constants ${\cal Q}_c$ are also indicated on the plots.

\section{Collapse}

We collapsed all the experimental data in  Fig. 3c in the main text. To collapse a $1/Q_g$ vs. $p$ data set, we divided each data point in the set by the factor ${\frac{S_r}{m_r} \sqrt {\frac{{\mu_g \rho_g \tau}}{{2\left(\omega_0 \tau+ \frac {\lambda} {l_x}\right)}}} {\cal Q}_{p}}$. The pressure-dependent quantities in this factor were properly evaluated. Once divided, each data point was plotted against its $\omega_0 \tau +\frac {\lambda} {l_x}$ obtained for the same pressure. We repeated this procedure for all the data sets.

A similar procedure was used for the cylinder fits as described in the main text. In this case, each $1/Q_g$ data point was divided by the pressure-dependent factor $T_0 {\cal Q}_c$. The divided data point was plotted against the ${\rm Re}_{\omega}$ value that was evaluated at the same pressure.

\section{Outlook}

It may not be far fetched to expect a similar scaling of a turbulent flow, because an expansion comparable to that in Eq.~(\ref{eq:stress_general})  has been shown to be applicable to the statistical theory of turbulence \cite{Chen}.  Furthermore, in the next order in the Chapman-Enskog expansion, thermal (energy) gradients are generated along with velocity derivatives. Large thermal gradients may  lead to a similar breakdown of Newtonian heat transfer (diffusion approximation)  with various interesting effects in close proximity to solid surfaces \cite{thermal, Gang_Chen}.

\begin{figure*}
\centering
\includegraphics[width=6.75in]{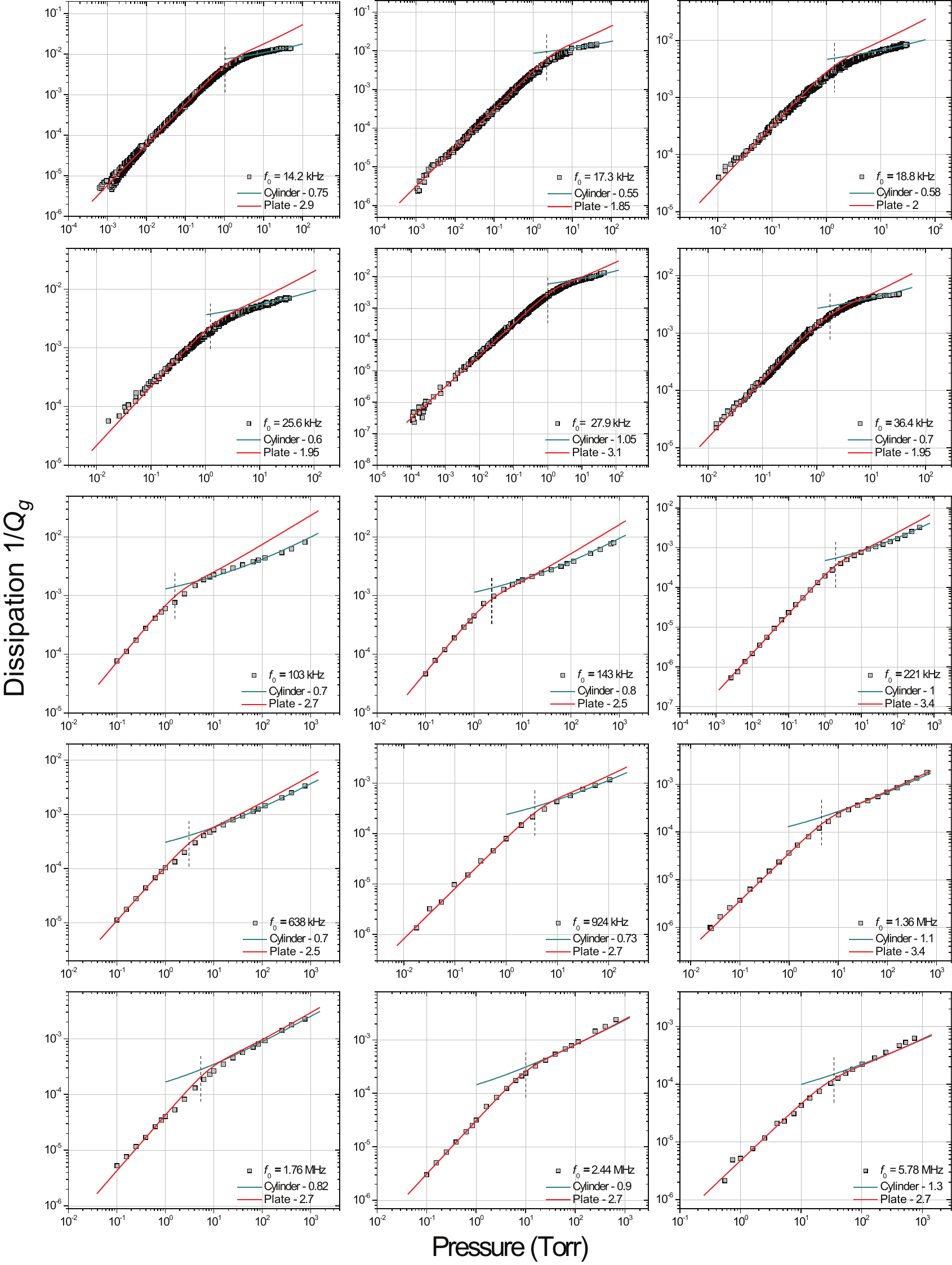}
\caption{Dissipation  of microcantilevers as a function of pressure  in N$_2$. The red line is the plate solution; the blue line is the cylinder solution. The dashed black vertical line shows the transition pressure $p_c$.}
\label{N2silicon}
\end{figure*}

\begin{figure*}
\centering
\includegraphics[width=6.75in]{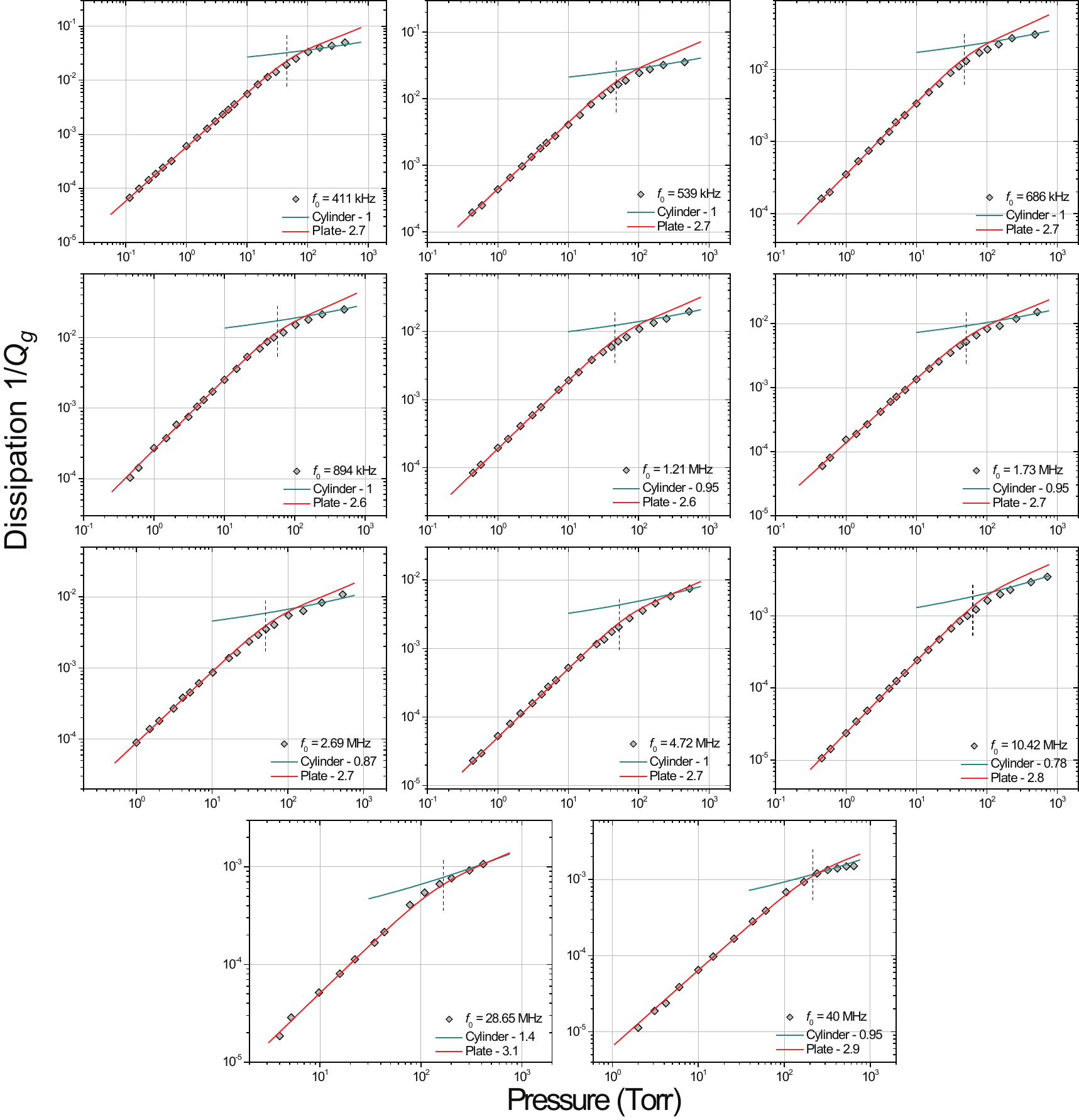}
\caption{Dissipation  of single-crystal diamond nanocantilevers  in N$_2$. Data for the 411 kHz, 686 kHz, and 40 MHz cantilevers were published in \cite{Kara} without the theoretical fits. The red line is the plate solution; the blue line is the cylinder solution. The dashed black vertical line shows the transition pressure $p_c$.}
\label{N2diamond}
\end{figure*}

\begin{figure*}
\centering
\includegraphics[width=6.75in]{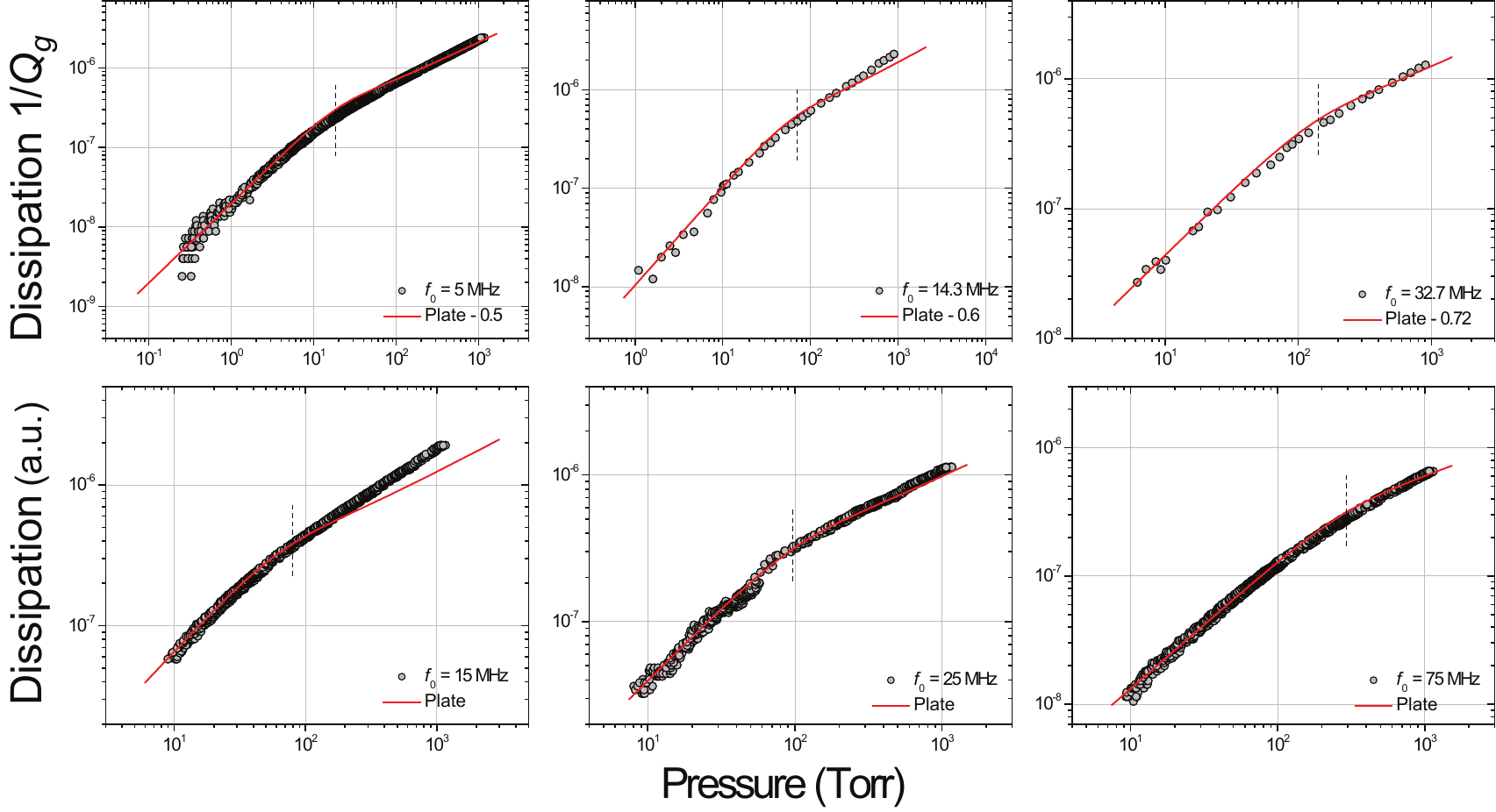}
\caption{Dissipation of  quartz crystal resonators as a function of pressure in N$_2$. The 14.3 MHz and the 32.7 MHz data  were published in \cite{EkinciPRL}. The red line is the plate solution. The dashed black vertical line shows the transition pressure $p_c$. Some data are left in arbitrary units; here, the fits were multiplied by arbitrary constants.}
\label{N2quartz}
\end{figure*}

\begin{figure*}
\centering
\includegraphics[width=6.75in]{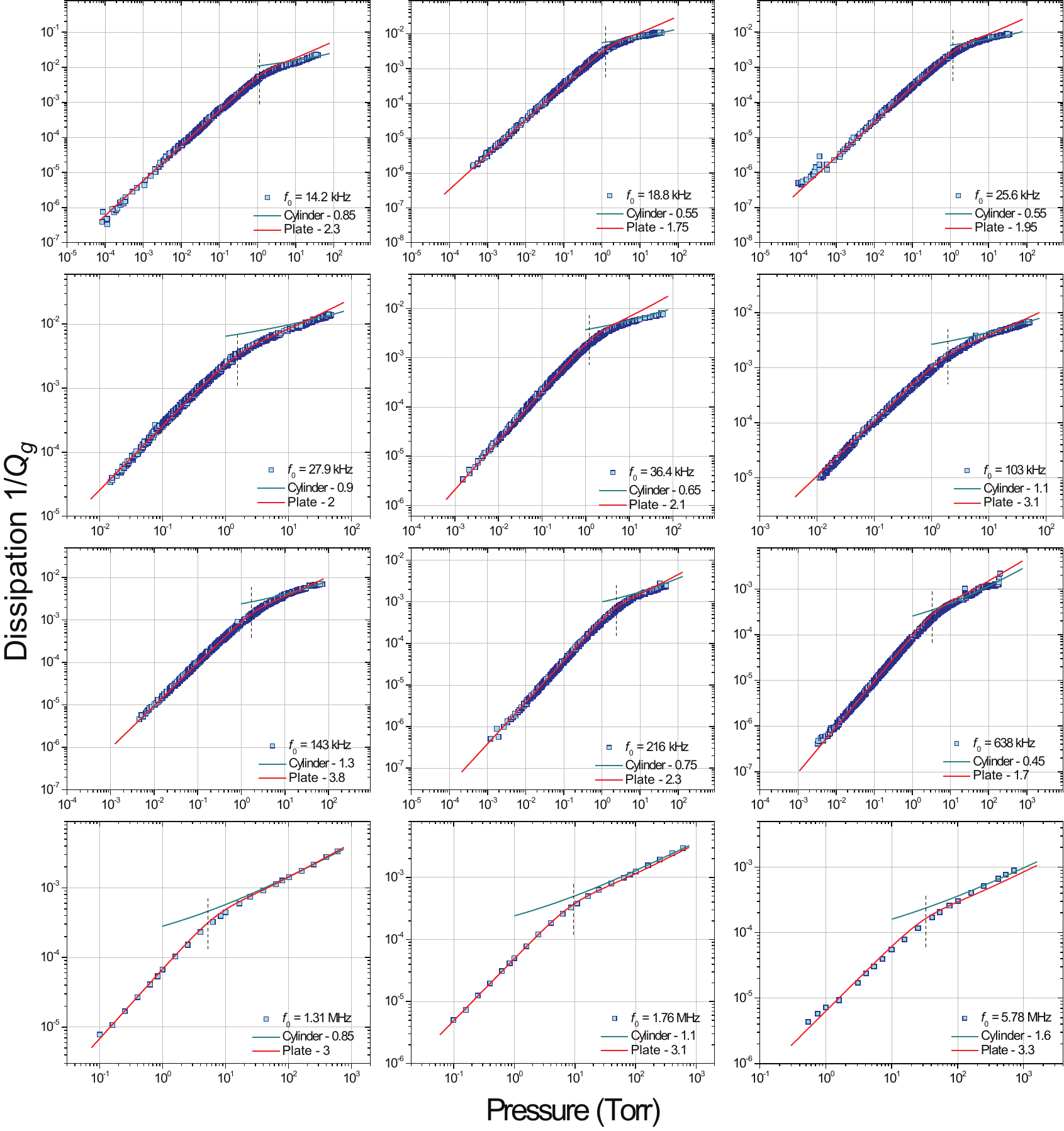}
\caption{Dissipation  of microcantilevers as a function of pressure  in Ar. The red line is the plate solution; the blue line is the cylinder solution. The dashed black vertical line shows the transition pressure $p_c$.}
\label{Arsilicon}
\end{figure*}

\begin{figure*}
\centering
\includegraphics[width=6.75in]{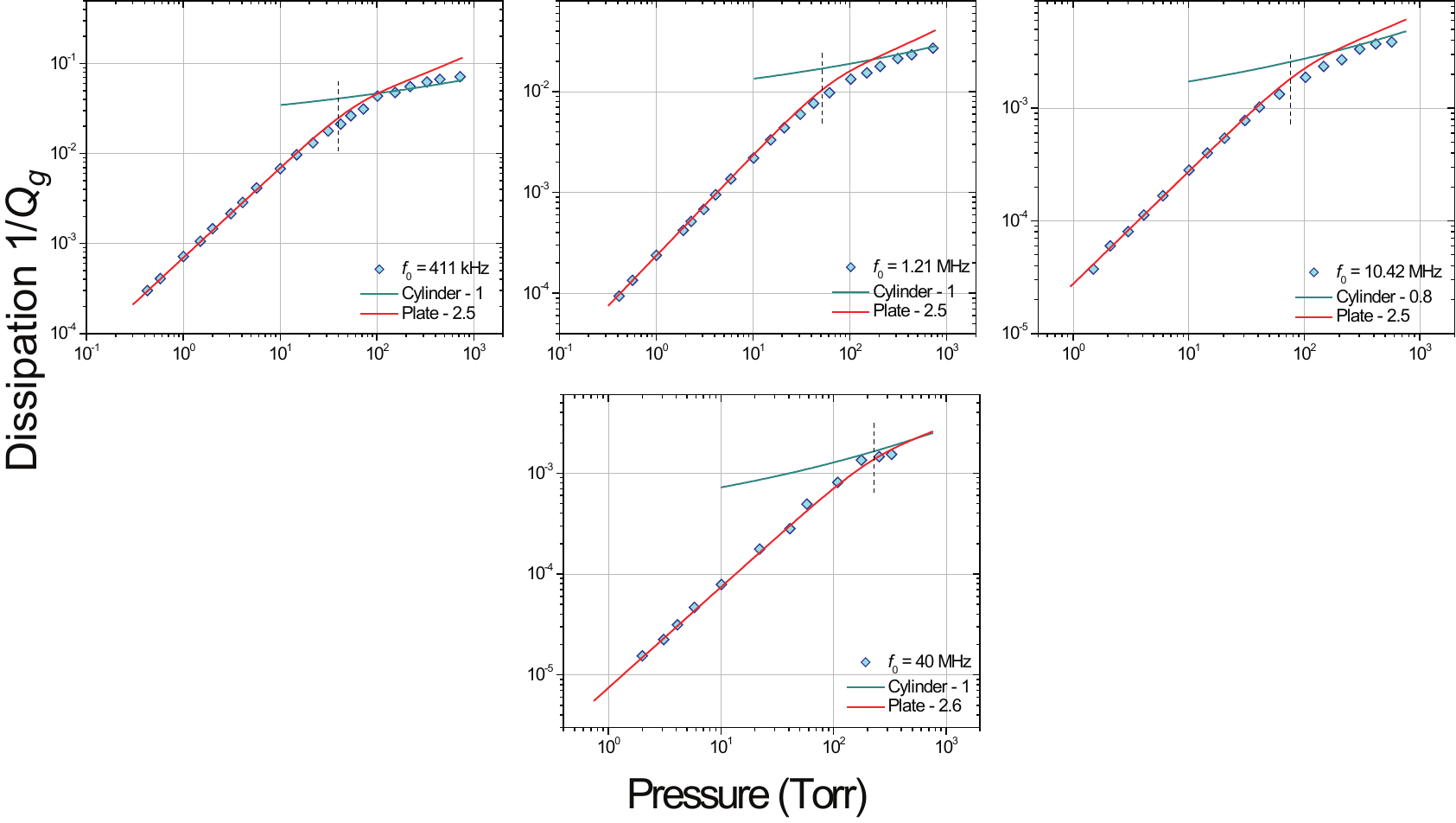}
\caption{Dissipation  of single-crystal diamond nanocantilevers  in Ar. Data for the 1.2 MHz cantilever were published in \cite{Kara} without the theoretical fits. The red line is the plate solution; the blue line is the cylinder solution. The dashed black vertical line shows the transition pressure $p_c$. }
\label{Ardiamond}
\end{figure*}

\begin{figure*}
\centering
\includegraphics[width=6.75in]{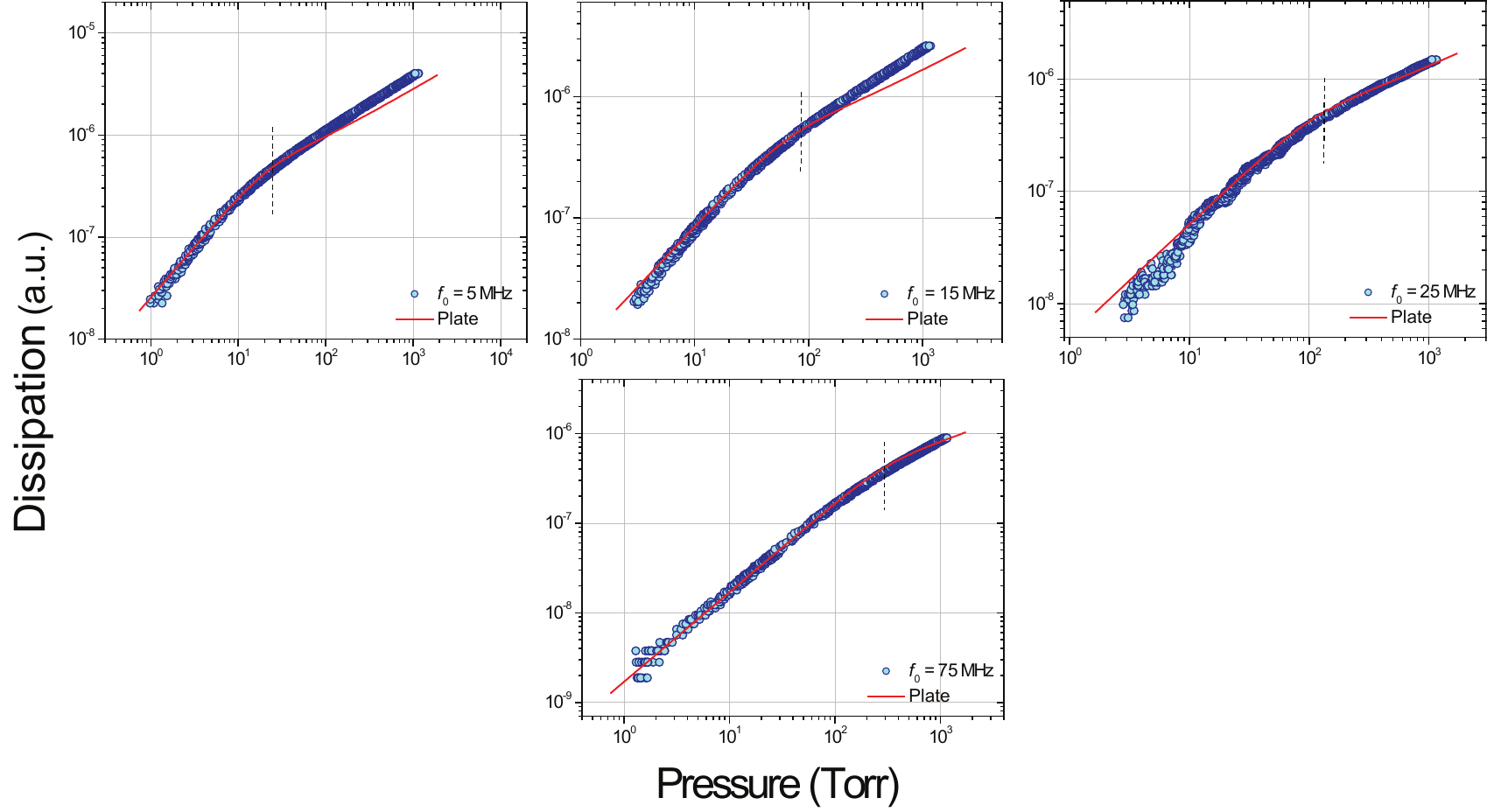}
\caption{Dissipation of quartz crystal resonators as a function of pressure in Ar. The red line is the plate solution. The dashed black vertical line shows the transition pressure $p_c$. Some data are left in arbitrary units; here, the fits were multiplied by arbitrary constants.}
\label{Arquartz}
\end{figure*}

\begin{figure*}
\centering
\includegraphics[width=6.75in]{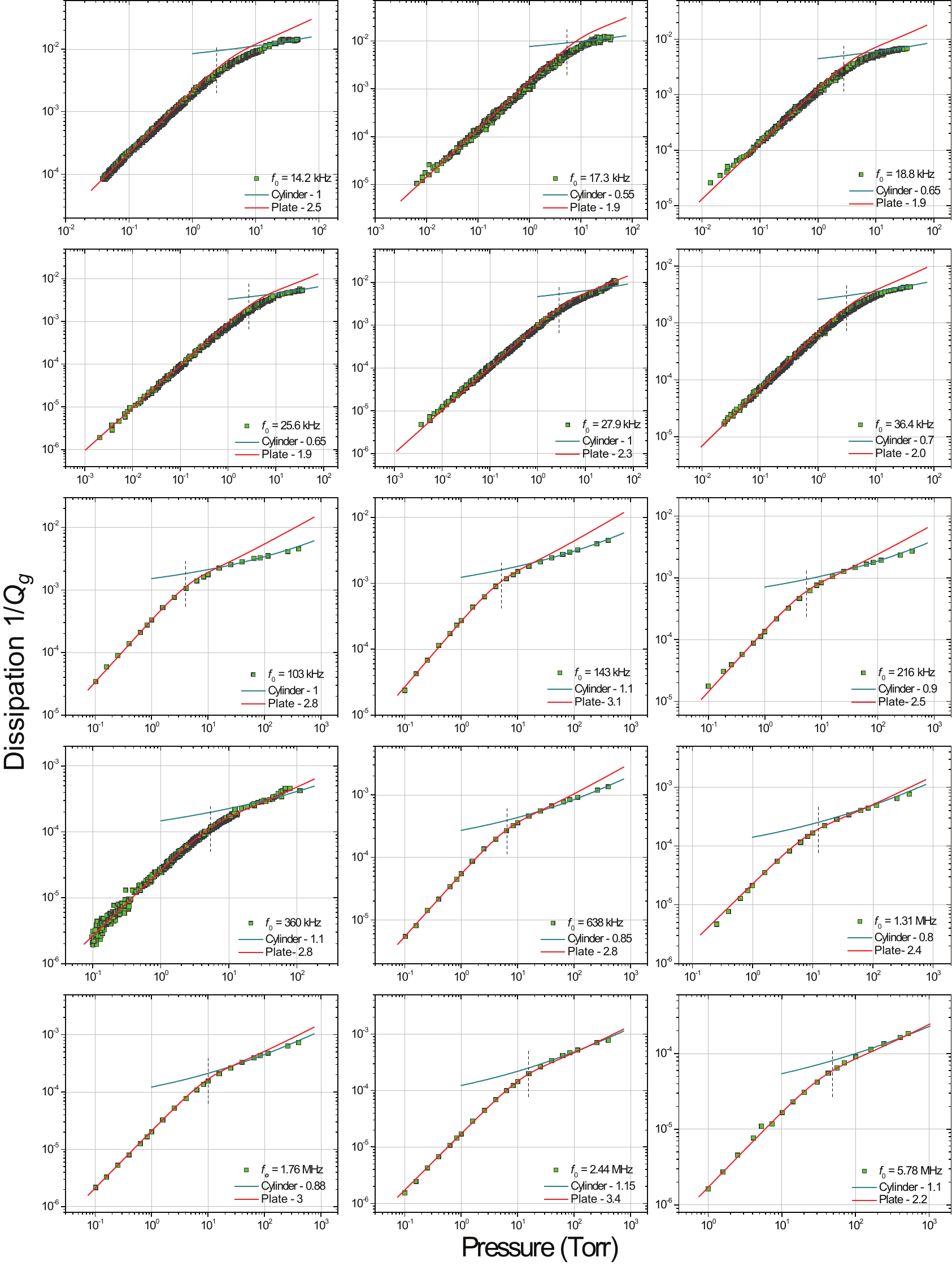}
\caption{Dissipation  of microcantilevers as a function of pressure  in He. The red line is the plate solution; the blue line is the cylinder solution. The dashed black vertical line shows the transition pressure $p_c$. }
\label{Hesilicon}
\end{figure*}

\begin{figure*}
\centering
\includegraphics[width=6.75in]{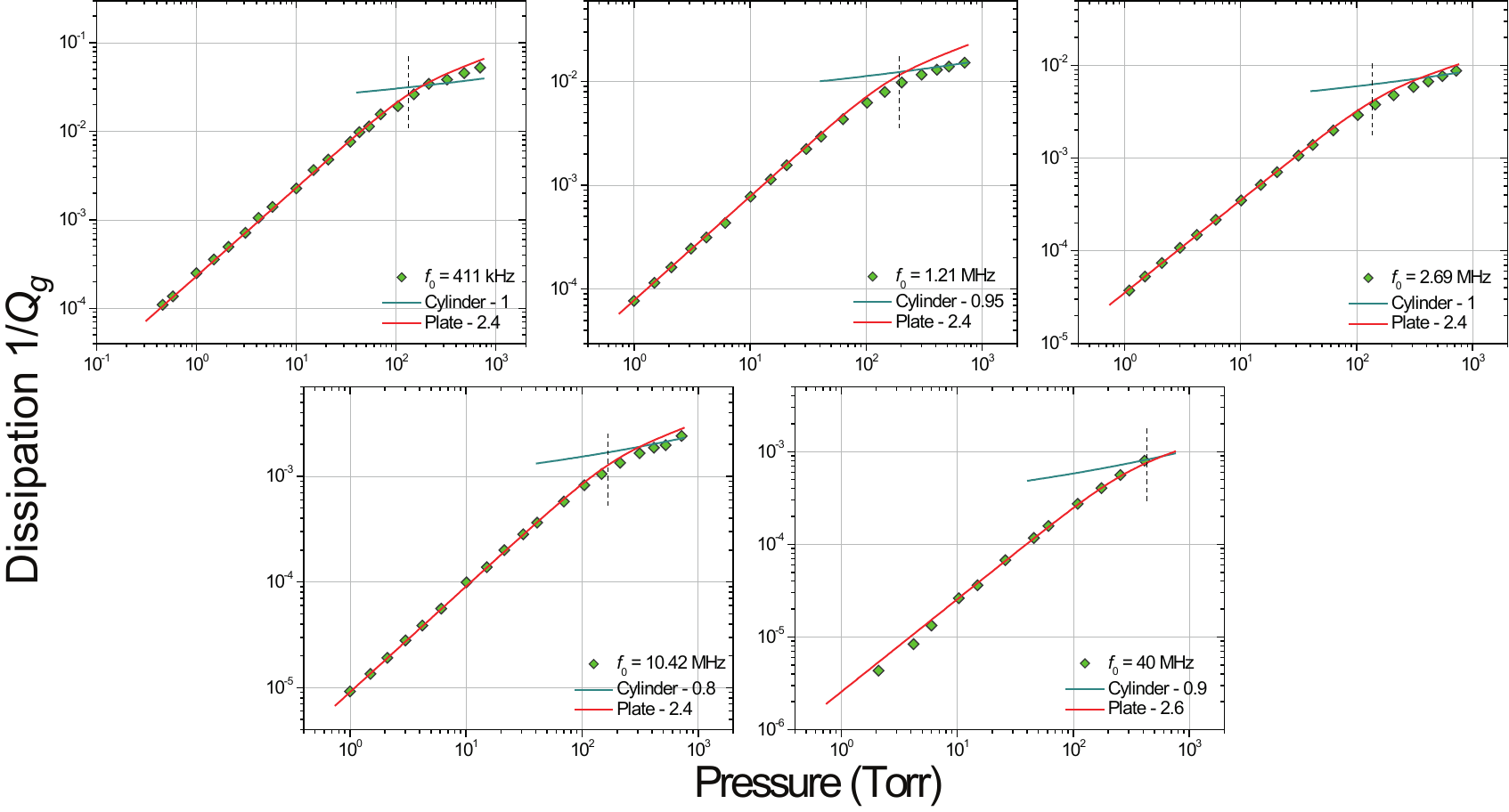}
\caption{Dissipation  of single-crystal diamond nanocantilevers  in He. Data for the 1.2 MHz cantilever were published in \cite{Kara} without the theoretical fits. The red line is the plate solution; the blue line is the cylinder solution. The dashed black vertical line shows the transition pressure $p_c$. }
\label{Hediamond}
\end{figure*}

\begin{figure*}
\centering
\includegraphics[width=6.75in]{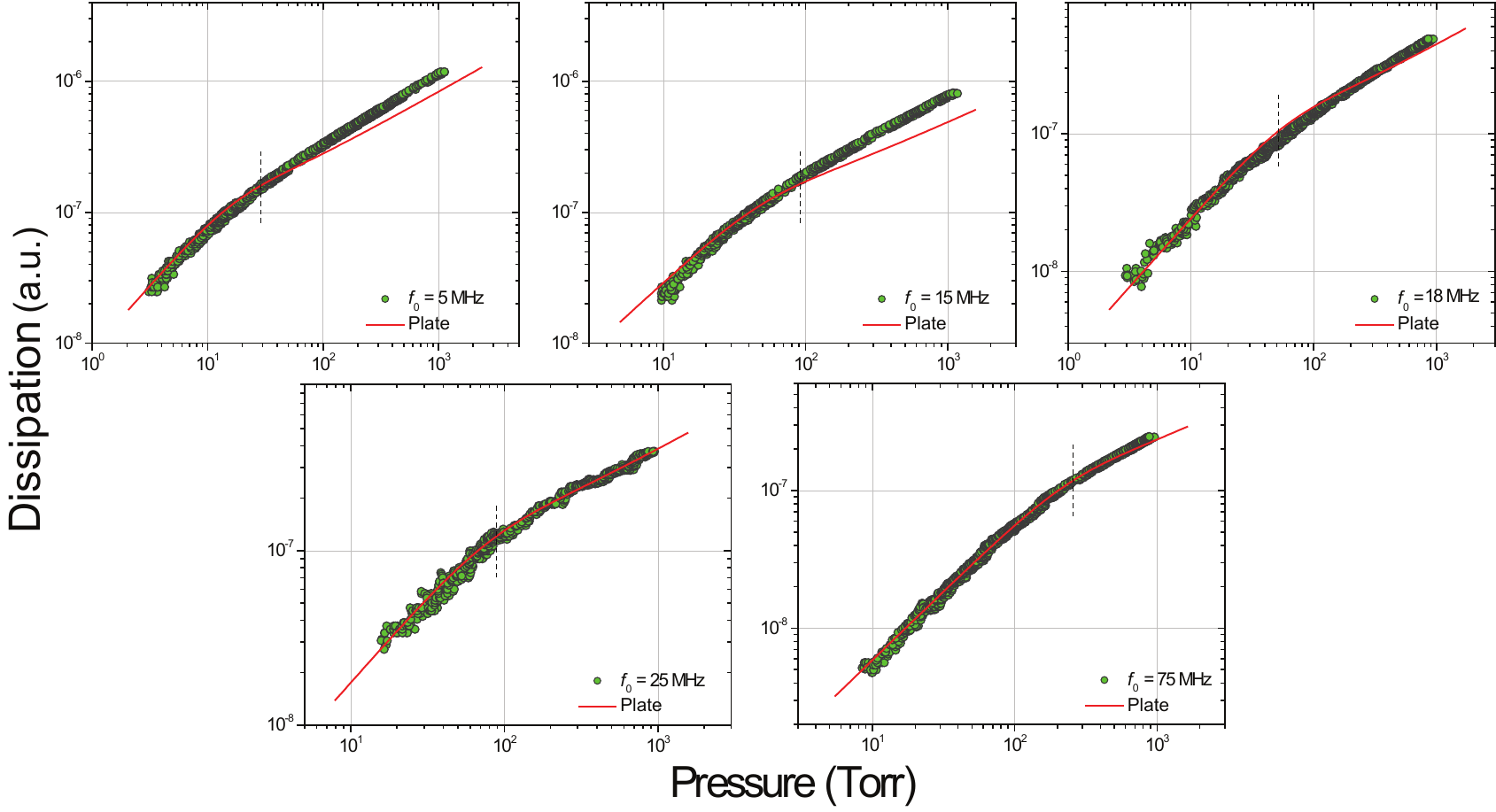}
\caption{Dissipation  of  quartz crystal resonators as a function of pressure in He. The dashed black line shows the transition pressure. The data are left in arbitrary units; here, the fits were multiplied by arbitrary constants.}
\label{Hequartz}
\end{figure*}